\documentclass[a4paper,11pt]{article}
\pdfoutput=1 

\usepackage{jheppub}
\usepackage{amsmath}
\usepackage{amssymb}
\usepackage{epsfig}
\usepackage{subfigure}

\newcommand{\dd}{{\rm d}}
\newcommand{\X}{{\rm X}}
\newcommand{\eq}[1]{(\ref{#1})}
\def\cO#1{{{\cal{O}}}\left(#1\right)}

\newcommand{\vac}{\ensuremath{{\rm vac}}}

\newcommand{\EQvac}{\ensuremath{{E}^\vac_Q}}
\newcommand{\EGvac}{\ensuremath{{E}^\vac_\gamma}}

\newcommand{\pQvac}{\ensuremath{{p}^\vac_Q}}
\newcommand{\pGvac}{\ensuremath{{p}^\vac_\gamma}}
\newcommand{\ptQvac}{\ensuremath{{p}^\vac_{T Q}}}
\newcommand{\ptGvac}{\ensuremath{{p}^\vac_{T \gamma}}}
\newcommand{\yQvac}{\ensuremath{{y}^\vac_{Q}}}
\newcommand{\yGvac}{\ensuremath{{y}^\vac_{\gamma}}}

\newcommand{\EQmed}{\ensuremath{{E}_Q}}
\newcommand{\EGmed}{\ensuremath{{E}_\gamma}}

\newcommand{\pQmed}{\ensuremath{{p}_Q}}

\newcommand{\ptQmed}{\ensuremath{{p}_{T Q}}}
\newcommand{\ptGmed}{\ensuremath{{p}_{T \gamma}}}
\newcommand{\yQmed}{\ensuremath{{y}_{Q}}}

\newcommand{\Etwovac}{\ensuremath{{E}^\vac_2}}
\newcommand{\pttwovac}{\ensuremath{{p}^\vac_{T 2}}}

\newcommand{\ptwomed}{\ensuremath{{p}_2}}
\newcommand{\Etwomed}{\ensuremath{{E}_2}}

\newcommand{\pttwomed}{\ensuremath{{p}_{T 2}}}
\newcommand{\ytwomed}{\ensuremath{{y}_{2}}}
\newcommand{\ytwovac}{\ensuremath{{y}^\vac_{2}}}

\newcommand{\DeltaE}{\ensuremath{\epsilon}}
\newcommand{\DeltaET}{\ensuremath{\epsilon}_T}
\newcommand{\DeltaEguess}{\ensuremath{\tilde{\epsilon}}}
\newcommand{\DeltaETguess}{\ensuremath{\tilde{\epsilon}}_T}

\newcommand{\qtvac}{\ensuremath{q_T^{\rm vac}}}
\newcommand{\qtmed}{\ensuremath{q_T^{\rm med}}}

\newcommand{\zGQ}{z_{\gamma Q}}

\newcommand{\p}{{\rm p}}
\newcommand{\A}{{\rm A}}

\newcommand{\raa}{R_{\rm AA}}

\newcommand{\pt}{p_{_T}}
\newcommand{\qt}{q_{_T}}

\def\cO#1{{{\cal{O}}}\left(#1\right)}
\newcommand{\sqrtsnn}{\sqrt{s_{_{\rm NN}}}}

\newcommand{\ptgam}{\ensuremath{{p_{T\gamma}}}}
\newcommand{\ptQ}{\ensuremath{{p_{TQ}}}}

\newcommand{\GeV}{ {\rm GeV}}

\def\msbar{\ensuremath{{\rm{\overline{MS}}}}}

\newcommand{\mur}{\ensuremath{\mu_R}}
\newcommand{\mui}{\ensuremath{\mu_F}}

\title{Prompt photon in association with a heavy-quark jet in Pb--Pb collisions at the LHC}

\author[a]{Tzvetalina\ Stavreva,}
\author[b]{Fran\c{c}ois\ Arleo,}
\author[a]{Ingo\ Schienbein}

\affiliation[a]{
Laboratoire de Physique Subatomique et de Cosmologie, UJF, CNRS/IN2P3,
\\
INPG, 53 avenue des Martyrs, 38026 Grenoble, France}
\affiliation[b]{Laboratoire d'Annecy-le-Vieux de Physique Th\'eorique (LAPTh)\\ UMR5108, 
Universit\'e de Savoie, CNRS, BP 110, 74941 Annecy-le-Vieux cedex, France}

\emailAdd{stavreva@lpsc.in2p3.fr}
\emailAdd{arleo@lapp.in2p3.fr}
\emailAdd{schien@lpsc.in2p3.fr}

\abstract{
We present a phenomenological study of the associated production of a
prompt photon and a heavy quark jet (charm or bottom) in Pb--Pb collisions at the LHC.
This channel allows for estimating the amount of energy loss
experienced by the charm and bottom quarks 
propagating in the dense QCD medium produced in those collisions.
Calculations are carried out at next-to-leading order (NLO) accuracy using the BDMPS-Z heavy-quark quenching weights.
The quenching of the single heavy-quark jet spectrum reflects fairly
the hierarchy in the heavy quark energy loss assumed in the perturbative calculation. 
On the contrary, the single photon spectrum in heavy-ion collisions is
only modified at low momenta, for which less heavy-quark jets pass the kinematic cuts.
On top of single particle spectra, the two-particle final state provides a range of observables (photon-jet pair momentum, jet
asymmetry, among others) which are studied
in detail. 
The comparison of the photon-jet pair momentum, from p--p to Pb--Pb
collisions, is sensitive to the amount of energy lost by the
heavy-quarks and could therefore be used in order to better understand
parton energy loss processes in the heavy quark sector.
}


\preprint{LPSC-12-323, LAPTH-056/12}

\begin{document} 
\maketitle
\flushbottom

\section{Introduction}

Spectacular experimental results on large $\pt$ particle production
have been reported in Pb--Pb collisions at the LHC by the ALICE,
ATLAS, and CMS experiments. 
The quenching of light hadrons (inclusive charged hadrons, pions, etc.) has
been observed over a very large range of transverse momenta~\cite{Aamodt:2010jd,ATLAS-CONF-2011-079,CMS:2012aa}, soon
followed by that of charmed mesons such as $D$ or $D^\star$~\cite{ConesadelValle:2012vp,ATLAS-CONF-2012-050}. 
Apart from large-$\pt$ hadron production, the reconstruction of jets
in heavy-ion collisions also allowed for the observation of
significant jet asymmetries of jet--jet~\cite{Collaboration:2010bu,Chatrchyan:2011sx}, photon--jet~\cite{Chatrchyan:2012gt} and Z--jet~\cite{ATLAS-CONF-2012-119} correlations, as well as measurements of 
hadron momentum spectra (the so-called fragmentation functions) inside jets~\cite{Chatrchyan:2012gw,ATLAS-CONF-2012-115}.

Clearly, these results are qualitatively consistent with parton energy loss processes as hard quarks and gluons propagate through the dense QCD medium produced in those collisions.
 On the more quantitative side many questions still remain unanswered, among which the issue of (radiative) energy loss in the heavy quark sector. 
 Because the heavy quark mass acts as a collinear cut-off in the medium-induced gluon radiation, the following hierarchy $\DeltaE_q > \DeltaE_c > \DeltaE_b$ of the typical energy loss of light, charm and bottom quarks has been proposed by Dokshitzer and Kharzeev~\cite{Dokshitzer:2001zm} and later checked by Armesto, Salgado and Wiedemann in the BDMPS-Z energy loss framework~\cite{Armesto:2003jh}.
Experimentally, however, the slight differences in the quenching of $D$ mesons with respect to that of pions might nevertheless be understood as coming from the different color charge of the propagating parton, essentially heavy quarks ($C_F=4/3$) and gluons ($C_A=3$) for $D$ mesons and pions, respectively. As of today, there is no clear and indisputable sign on possible differences between the energy loss of light and heavy quarks. On the theoretical side, moreover, the above ``standard'' hierarchy has also been recently questioned by Aurenche and Zakharov in Ref.~\cite{Aurenche:2009dj} due to formation time effects.
It is therefore crucial to identify observables at the LHC which could help clarify this issue. 
In this paper, we argue that measuring prompt photon production in association with a heavy quark tagged jet (denoted as $\gamma+Q$ in the following) in heavy-ion collisions might shed light on the mass dependence of the radiative parton energy loss mechanism.

The production of $\gamma+Q$ , first measured in p--$\bar{\rm  p}$
collisions at the Tevatron by the 
D0 and CDF experiments~\cite{Abazov:2009de,Aaltonen:2009wc,Abazov:2012ea,D0:2012gw}, 
is a rich and versatile process in various hadronic collisions:
\begin{itemize}
\item In p--p and p--$\bar{{\rm p}}$ collisions, first of all, $\gamma+Q$ production offers sensitive checks 
of perturbative QCD (pQCD) and might serve as a probe of intrinsic
heavy quark distributions inside the proton~\cite{Stavreva:2009vi};
\item In  p--A collisions (e.g. at RHIC and soon at the LHC), this process
can be used to constrain the gluon parton distribution function in nuclei (nPDF)~\cite{Stavreva:2010mw}, which 
is pretty much unconstrained at small values of $x$.
  One should underline that knowing precisely the nuclear PDFs is a prerequisite 
  in order to obtain reliable predictions in  heavy-ion  collisions;
\item In A--A collisions, finally, the study of $\gamma+Q$ provides an ideal tool for investigating the energy lost by
  heavy quarks in the hot QCD medium produced in those collisions.  Being an electromagnetic probe, the
  photon produced directly in the hard process is expected to traverse
  the medium unaffected. Its momentum can therefore serve as a proxy
  for the initial momentum of the heavy quark  propagating through the
  dense medium and eventually fragmenting into the heavy quark jet. 
The imbalance between the prompt photon and the heavy quark jet
momentum going from p--p to A--A collisions might thus reflect the amount of
energy loss experienced by the heavy quark. 
Furthermore, the  comparison between $\gamma + c$ and $\gamma + b$
production would provide access to the mass hierarchy of parton energy loss.
\end{itemize}

It is the purpose of this paper to perform an exploratory study
of heavy quark energy loss from a next-to-leading order (NLO) QCD analysis of $\gamma+Q$ production in heavy-ion collisions.\footnote{A related but different analysis of prompt photon and heavy quark \emph{hadrons} has been performed previously at LO accuracy in the fixed-flavor number scheme~\cite{Kang:2011rt}. Earlier studies investigating photon correlations with inclusive hadrons have been performed at NLO, see e.g.~\cite{Arleo:2004xj,Zhang:2009rn}. More recently $\gamma$ correlations with inclusive jets were also investigated~\cite{Dai:2012am}.}
In addition to the photon and the heavy quark $\pt$ spectra,
we study the distributions in different kinematic variables 
($q_T$, $A_J$, $z$)
which prove useful in order to characterize the amount of heavy quark energy loss in the medium.
Note that in these latter cases, it is mandatory to perform the theoretical
calculation
of the $\gamma+Q$ production process
 at NLO in order to obtain reliable results since these
variables exhibit a singular behavior at LO accuracy.
However even for the more inclusive observables such as the $p_T$-distributions of
either the photon or the heavy quark, which are not singular at LO, the
NLO calculation is important for obtaining the right magnitude of the cross
section.
As we shall see in Section~\ref{sec:eloss}, however, the treatment of parton energy loss processes in heavy-ion collisions is performed at leading order only.

This paper is organized as follows.
In Section~\ref{sec:framework} we describe our theoretical framework
for the $\gamma + Q$ production including the
effects of the medium formed during the A--A collisions.
Our numerical results are presented in Section~\ref{sec:numerics}.
Finally, in Section~\ref{sec:conclusions} we present our conclusions
and give an outlook on possible future improvements of the analysis. 

\section{Theoretical framework} 
\label{sec:framework}

\subsection{$\gamma + Q$ production in p--p collisions}
\label{sec:vac}

Using the standard pQCD collinear factorization framework the cross section for the production 
of a photon and a heavy quark in p--p collisions is given by
\begin{equation}
\dd\sigma = \sum_{i,j,k}f_{i}\otimes f_{j}\otimes \dd\hat \sigma(i j
\rightarrow k Q) \otimes D_k^\gamma \, ,
\label{Eq:fac}
\end{equation}
where the sum over all possible partonic subprocesses is performed and $f_i$ stands for the parton 
distribution function in the proton. 
Furthermore,  $D_k^\gamma$ describes the fragmentation of the final state parton $k$ into the observed photon
and  the direct contribution ($k=\gamma$) is included via $D_\gamma^\gamma(z)=\delta(1-z)$.

The contributing partonic subprocesses depend on the heavy flavor scheme. 
The present calculation is performed in the variable flavor number scheme (VFNS), 
which includes heavy quark PDFs for factorization scales greater than
the threshold given by the heavy quark mass, $\mu_{\rm F} \ge m_Q$. 
 Furthermore, heavy quark mass terms $m_Q^2/\mu^2$ have been neglected
 in the perturbative calculation of the partonic cross sections 
 where the factorization/renormalization scale $\mu$ is identified with a typical hard scale of the process, that is
 the transverse momentum of the photon or the jet.\footnote{Note that in the case of a gluon 
 splitting into a collinear $Q \bar Q$ pair, the divergency is avoided by imposing a cut on 
 the invariant mass such that $m_{Q\bar Q}>2 m_Q$~\cite{Stavreva:2009vi}.}
The range of validity of the calculation is thus restricted to
 momenta $\mu=\cO{\pt}$ larger than a few times the heavy quark mass.
Note that in heavy-ion collisions the factorization ansatz in Eq.~\eqref{Eq:fac} is a working assumption.

\begin{table}[h]
\begin{center}
\begin{tabular}{cc}
\hline
\hline
$gg\rightarrow \gamma Q\bar Q$ &
$gQ\rightarrow \gamma gQ$\\
$Qq\rightarrow \gamma qQ$ &
$Q\bar q\rightarrow \gamma \bar qQ$ \\
$Q\bar Q\rightarrow \gamma Q\bar Q$&
$QQ\rightarrow \gamma QQ$ \\
$q\bar q\rightarrow \gamma Q\bar Q$\\
\hline
\hline
\end{tabular}
\caption{\label{table:NLO}List of all direct $2\rightarrow 3$ NLO hard-scattering 
subprocesses.}
\end{center}
\end{table}

At leading-order accuracy, ${\cal O}(\alpha\alpha_s)$,  the production of a {\it direct} photon with a
heavy quark jet only arises from the $g Q \to \gamma Q$ Compton
scattering subprocess at the hard-scattering level.
At next-to-leading order, however, the number of contributing subprocesses increases to seven (see Table~\ref{table:NLO}).  
On top of these direct photon subprocesses, the contributions coming from the fragmentation  of a parton into a photon 
are also included consistently at NLO accuracy.
It should however be mentioned that isolation requirements --~used experimentally in order to 
minimize background coming from hadron decays~-- greatly decrease these fragmentation contributions. 
For further details on the theoretical calculations, the 
reader may refer to~\cite{Stavreva:2009vi,Stavreva:PhD}. 

\subsection{$\gamma + Q$ production in heavy-ion collisions}
\label{sec:eloss}

We discuss in this section the calculation of $\gamma+Q$ production in
heavy-ion collisions, in which the final-state partons propagating
through the dense QCD medium are expected to lose energy through 
medium-induced gluon radiation, or \emph{radiative} energy loss
processes.\footnote{We do not consider collisional energy loss
processes in this study, as they are expected to be less important (in comparison to radiative losses) at large momenta.}

Let us first denote by $\pQvac$ and $\pGvac$ the 4-momenta of the 
heavy quark and the photon, respectively, in p--p collisions. 
We express these momenta  in terms of their transverse
momentum $p_T^\vac$ and  rapidity $y^\vac$:
\begin{eqnarray}
\pQvac &=& \ptQvac (\cosh \yQvac, \vec{e}_{TQ}, \sinh \yQvac) \ ,\\
\pGvac &=& \ptGvac (\cosh \yGvac, \vec{e}_{T\gamma}, \sinh \yGvac)\ .
\end{eqnarray}
Because of energy loss processes, in heavy-ion collisions 
the energy of the heavy quark
is shifted with respect to that in p--p collisions, $\EQmed = \EQvac - \DeltaE$,
where $\DeltaE$ is the amount of energy lost by the heavy quark
while traversing the medium.\footnote{Strictly speaking, $\DeltaE$ should be understood as the energy radiated \emph{outside} the jet cone. We therefore implicitly assume in the present study that most of the medium-induced gluon radiation occurs at large angles.}
In the present exploratory study, we assume that the photon is not affected by the medium,
 $\EGmed = \EGvac$, and moreover  that the heavy quark
does not change its direction while propagating through the medium, $\yQmed=\yQvac$.
Therefore, the heavy quark momentum in the medium is given by
\begin{equation}\label{eq:shiftQ}
\pQmed = \ptQmed (\cosh \yQmed, \vec{e}_{TQ}, \sinh \yQmed) 
=[\ptQvac - \DeltaE/\cosh \yQvac] (\cosh \yQvac, \vec{e}_{TQ}, \sinh \yQvac)  \, ,
\end{equation}
i.e., 
\begin{equation}
\ptQmed = \ptQvac - \DeltaET\, , \, \DeltaET = \DeltaE/\cosh \yQvac\, .
\end{equation}

As discussed in the previous section, beyond the leading order a second parton (labeled ``2'') is produced in real $2\to3$ subprocesses 
(see Table~\ref{table:NLO}).
The emission of this extra parton occurs within the short-distance time scale $\cO{Q^{-1}} \ll 1$~fm (where $Q \gg \Lambda_{\rm QCD}$ is the scale of the hard process), therefore well \emph{before} the medium is produced. As a consequence, parton 2 also experiences medium-induced energy loss; 
its energy is thus shifted, $\Etwomed = \Etwovac - \DeltaE^\prime$, leading to the following expression
for its momentum in medium similar to (\ref{eq:shiftQ}),
\begin{eqnarray}
\label{eq:shift2}
\ptwomed &=& [\pttwovac - \DeltaE^\prime/\cosh \ytwovac] (\cosh \ytwovac, \vec{e}_{T2}, \sinh \ytwovac)\, , \\
\pttwomed &=& \pttwovac -  \DeltaET^\prime\, , \,  \DeltaET^\prime =  \DeltaE^\prime/\cosh \ytwovac\, , \, \ytwomed = \ytwovac \, , \, \nonumber
\end{eqnarray}
after losing the energy $\DeltaE^\prime$ in the medium.

Although the dynamics of $\gamma+Q$ production is performed at NLO in order to get meaningful results for the 2-particle distributions, we stress that energy loss processes are treated at leading order accuracy. In particular
the momenta of the two final-state partons produced in $2\to3$ processes are shifted \emph{independently}.
In doing so 
 we neglect the possible interference effects in the medium-induced gluon radiation off two partons.
 This question has been addressed in a recent series of papers, see for example~\cite{MehtarTani:2010ma} as well as~\cite{Armesto:2011ir} for the treatment involving massive partons.
Completing the NLO treatment of this process by including the energy loss  coherence effects is of great interest, and will be the aim of a future study. 
There, also the issue of  
subsequent (soft and collinear) radiation can be addressed, these occuring on longer time scales and possibly after the radiating parton has escaped the medium (see~\cite{CasalderreySolana:2011gx});  such radiation,
which might be more sensitive to the above coherence effects,
 is not taken into account in the present fixed-order calculation scheme 
but could rather be studied through Monte Carlo parton showers. A recent discussion on coherence effects on the jet evolution in a medium can be found in~\cite{CasalderreySolana:2012ef}.

Note that the energy loss of the extra parton affects the production of $\gamma+Q$ events as long as it is recombined with the heavy quark $Q$ to form the heavy-flavor jet through a jet reconstruction algorithm\footnote{
For the jet reconstruction we combine the heavy quark and parton 2 in a jet if they are
within a cone of radius $R$ specified in Table~\protect\ref{tab:AA-cuts}.}, and not otherwise.

In order to compute the medium-modified $\gamma+Q$ production cross section in heavy-ion collisions, we use the following Monte Carlo procedure:
\begin{itemize}
\item[1.]  We obtain an event for the production  of a  $\gamma+ Q$  
in vacuum at a given heavy quark four momentum $\pQvac$ and 
photon four momentum, $\pGvac$.
\item[2.] 
For each partonic event (i.e. with one or two partons in the final state), the energy loss $\DeltaE$ (respectively, $\DeltaE^\prime$) of the heavy quark (respectively, the extra parton) is sampled according to a probability distribution 
or \emph{quenching weight}, ${\cal P}_i(\DeltaE)$, using an acceptance-rejection algorithm.
\item[3.] 
According to Eqs.\ \eqref{eq:shiftQ} and \eqref{eq:shift2},
 medium-modified four-momenta $p_Q$, $p_\gamma$ and $p_2$ are constructed.
The additional constraints, $\DeltaE < \EQvac$ and $\DeltaE^\prime < \Etwovac$, are moreover imposed.
\item[4.] 
Using the modified four-vectors we evaluate observables ($\ptQ, \ptgam$), as well
as the correlation variables discussed in Sec.~\ref{sec:observables}
in which we bin the events, providing the differential cross section for these observables.
\end{itemize}

\subsection{Quenching weights}
\label{sec:QW}

The quenching weights used in the present calculation are obtained using the Poisson approximation proposed in~\cite{Baier:2001yt},
\begin{equation}
\label{eq:Poisson}
{\cal P}_j(\DeltaE) = \sum^\infty_{n=0} \, \frac{1}{n!}\ \left[ \prod^n_{i=1} \, \int_0^{\DeltaE} \, \dd\omega_i \, {\frac{\dd{I}(\omega_i)}{\dd\omega}}\bigg|_j \right] 
\times  \, \delta \left(\DeltaE- \sum_{i=1}^n  \omega_i\right) \, \exp{ \left\{- \int_0^{\infty} \dd\omega \, \frac{\dd{I}}{\dd\omega} \right\} }\, ,
\end{equation}
where $j$ is the flavor of the propagating parton ($j=q,g,Q$) and $\dd{I}/\dd\omega$ the medium-induced gluon spectrum, which in the present work is determined from the perturbative BDMPS-Z framework~\cite{Wiedemann:2000za}.

The quenching weights ${\cal P}_j(\DeltaE)$ are a scaling function of $\DeltaE/\omega_c$, where $\omega_c\equiv1/2\ \hat{q}\ L^2$ is the typical scale for the energy loss process; the transport coefficient $\hat{q}$ measures the scattering power of the medium (momentum broadening per unit length) and $L$ is the medium length. The medium-induced gluon spectrum (and therefore the quenching weights) depends on the dimensionless parameter $R$, $R\equiv\omega_c\ L$, which arises from the kinematic constraint restricting the transverse momenta of the radiated gluons. The limit $R\to\infty$ of the spectrum corresponds to the ``thick medium'' limit of the BDMPS formulation~\cite{Baier:1996kr} in which the opacity, or the number of rescatterings  $n=L/\lambda\to\infty$.\footnote{Note that in this limit, the probability for no energy loss vanishes.}

In our numerical analysis we use the quenching weights provided in~\cite{Salgado:2003gb} 
for massless quarks and gluons and in~\cite{Armesto:2005iq} for heavy quarks. In the latter case, the weights also depend on the heavy quark mass through the dimensionless quantity $m/E$, where $E$ is the heavy quark energy in the medium rest frame. In this exploratory study, we will stick to the BDMPS framework, i.e. using the quenching weights computed in the multiple scattering scenario (as opposed to the $n=1$ opacity case) and taking the ``thick'' limit, $R\to\infty$.\footnote{For practical purposes, in order to allow for reasonable CPU time, we use $R=10^5$.  This seems to be a good approximation since the probability for no energy loss, $p_0$, is already below the few percent level for the $m/E$ values used in the calculation. }

Due to the heavy quark mass, the medium-induced radiation is suppressed at large gluon energies, see~\cite{Armesto:2003jh}. As a consequence, the typical energy loss in this calculation is substantially reduced when $m/E$ is large. This calculation confirms in particular the hierarchy, $\DeltaE_q > \DeltaE_c > \DeltaE_b$, conjectured in~\cite{Dokshitzer:2001zm} for heavy quark energy losses in QCD media.\footnote{At small values of the heavy quark mass, $m/E\sim 10^{-2}$ and intermediate values $R\sim 10^3$--$10^4$, the energy loss of massive quarks is actually \emph{larger} than in the massive case, questioning the hierarchy $\DeltaE_q > \DeltaE_c > \DeltaE_b$.
It is however not clear in~\cite{Armesto:2003jh} whether this observation has a well defined physical origin or is an artifact of the calculation.} As mentioned in the introduction, this hierarchy has recently been questioned by Aurenche and Zakharov in~\cite{Aurenche:2009dj} on the basis of formation time arguments. It is not the goal of the present paper to answer this question, nor is it to perform a comprehensive phenomenological study which would encompass various assumptions regarding heavy quark energy loss and a full treatment of the geometry and the dynamics of the produced medium. Rather we explore how the properties of the medium-induced gluon radiation off massive quarks translates into a variety of $\gamma+Q$ observables through an NLO calculation ---~within one well defined scheme for heavy quark energy loss.

In the next section, we present the different distributions which we consider in this study that we believe would best reflect the dynamics of massive quark energy loss processes in heavy-ion collisions.

\subsection{Observables}
\label{sec:observables}

The most inclusive observable which should be sensitive to the
energy loss is the total cross section in dependence of the
cut on the $\ptQ$ (or $\ptgam$).
In addition it is useful to study the inclusive transverse momentum 
distributions of the heavy quark and the photon, respectively.

On top of the photon and the heavy quark jet single $\pt$ spectra, the
two-particle final state further offers a wide range of observables
which might give a better access to the energy loss of the propagating
heavy quark. 
One such variable is the transverse momentum difference between the photon and the heavy quark jet, $\qt$, defined as follows:
\begin{equation}
 \qt = \ptGmed - \ptQmed \, .
 \label{eq:qT}
\end{equation}
At leading order and assuming that the photon is produced directly in
the hard process (i.e. not by fragmentation), the photon and the jet
momenta balance in the transverse plane. Therefore this variable
reduces at LO to $\qtvac =\ptGvac-\ptQvac=0$ in the vacuum (p--p
collisions) and $\qtmed
=\ptGmed-\ptQmed=\ptGvac-(\ptQvac-\DeltaET)=\DeltaET$ in the medium
(A--A collisions). 
Therefore, a non-zero cross section at a given $\qtmed \ne 0$
can be interpreted (at LO, direct) to be proportional to
the number of events 
where the heavy quark jet has suffered an energy loss $\DeltaET = \qtmed$ in the medium.
It is unfortunately not as simple
because the ``vacuum'' $\qt$ distribution in p--p collisions 
is non-zero at $\qt > 0$ as soon as fragmentation photons
or NLO contributions are taken into account
(the former being however suppressed from the use of isolation criteria).

Assuming the photon and the jet to be 
in different hemispheres
(see also the cuts in Table~\ref{tab:AA-cuts})
the momentum difference in A--A collisions becomes $\qtmed=\ptGvac-(\ptQvac-\DeltaET)=\qtvac+\DeltaET$.
In other words the $\qt$ spectrum in A--A collisions is shifted by $+\DeltaET$ as compared to p--p collisions, 
where $\DeltaET$ should be understood as a typical amount of (transverse) energy loss. 
The shift should therefore be more pronounced for light jets than for charm and bottom quark jets, 
because of the hierarchy used in the calculation.

Similar to the photon-jet pair momentum we shall investigate as well the imbalance $A_J$,
\begin{equation}
A_J= {{\ptGmed-\ptQmed}\over {\ptGmed+\ptQmed}}\, , 
\label{eq:AJ}
\end{equation}
 which has been measured for dijet \cite{Collaboration:2010bu,Chatrchyan:2011sx,Chatrchyan:2012nia}
 and photon-jet correlations \cite{Chatrchyan:2012gt} by ATLAS and CMS.
Finally the distribution in the momentum imbalance, 
\begin{equation}
\zGQ= -{ { \vec{p}_{T\gamma} . \vec{p}_{TQ} }\over {p_{T\gamma}^2}}
\end{equation}
is also considered. 
In the leading order kinematics (and assuming the photon is produced directly in the hard process), these two variables 
would reduce respectively to 
$A_J^{\rm med}={\DeltaET} / {(2\ptGvac-\DeltaET})$ 
and  
$\zGQ^{\rm med}=1-{{\DeltaET} / \ptGvac}$ 
in heavy-ion collisions. 
As compared to the $\qt$ variable, these are sensitive to the ratio of the energy lost over the 
parton energy (or, fractional energy loss), 
$\DeltaET/\ptGvac$, 
instead of the absolute magnitude of the energy loss, $\DeltaET$.

Generically, for all the above observables
quenching factors  $\raa$ are defined as
\begin{equation}
\raa = \frac{1}{N_{\rm coll}}\  \frac{\dd \sigma(\A + \A \to \gamma + Q)}{\dd \sigma(\p + \p \to \gamma + Q)}\, ,
\end{equation}
where $N_{\rm coll}$ stands for the number of binary nucleon--nucleon collisions ($N_{\rm coll}=A^2$ in minimum bias A--A collisions).
In particular, we consider the suppression factors for the
inclusive $\ptgam$ spectrum, $R_{AA}^\gamma(\ptgam)$, 
and for the inclusive $\ptQ$ spectrum, $R_{AA}^Q(\ptQ)$.

Some comments are in order:
(i) With this definition the ratios $R_{AA}^\gamma$ and $R_{AA}^Q$ would in
general be different from unity even in
the absence of cold and hot nuclear matter effects due to differences
between parton distributions inside protons and neutrons
as is the case for inclusive photon production (see e.g.~\cite{Arleo:2006xb}).
However, in our case the dominant initial state is $g+Q$ and these
parton distributions should be (to a very precise level) the same
inside protons and neutrons.
(ii) To construct this ratio experimentally, the p--p cross section which is
measured at a different center-of-mass energy has to be extrapolated
to the center-of-mass energy of the A--A run.
For this reason it is extremely valuable that the photon transverse
momentum spectrum which is insensitive to hot medium effects
(up to a sensitivity on the cut on the heavy quark jet energy)
can be used to calibrate the effects 
of the energy extrapolation and of cold nuclear matter.
(iii)
An advantage of considering
the double ratios $R_{AA}^{Q/\gamma}
= R_{AA}^Q / R_{AA}^\gamma$ is that uncertainties due to the choice of
scales and nuclear PDFs should largely cancel.
While the scale uncertainties already cancel to a good degree in the simple ratios
there remains a residual uncertainty due to the lack of knowledge of the
nuclear gluon distribution, see e.g., \cite{Stavreva:2010mw}. 
Hence, it will be crucial to use data from the
upcoming p--Pb run at the LHC at the beginning of 2013 to better constrain
the nuclear gluon distribution.\footnote{Alternatively, constraints on the nPDFs could 
possibly be obtained through \emph{inclusive} photon and weak boson production in A--A collisions.} 
Therefore, for the time being we refrain from using nuclear PDFs 
\cite{Eskola:2009uj,Hirai:2007sx,Schienbein:2009kk,deFlorian:2011fp} 
in this study where we focus on the effects of the heavy quark energy loss solely.


\section{Results}
\label{sec:numerics}

\subsection{Ingredients}
\label{sec:ingredients}

The present calculations have been carried out using the 
CTEQ6.6M PDFs inside a proton \cite{Nadolsky:2008zw} (along with the corresponding
strong coupling constant $\alpha_s^{\msbar,5}(M_Z)=0.118$ at
next-to-leading order) and the photon fragmentation functions of Bourhis, Fontannaz and Guillet \cite{Bourhis:1997yu}.
The renormalization, factorization and fragmentation scales have been set to 
$\mur=\mui=\mu_f=\ptgam$ 
and we have used $m_c=1.3\ \GeV$ and $m_b=4.2\ \GeV$ for the charm 
and bottom quark masses.
All cross sections (for both p--p and Pb--Pb collisions) have been
calculated at a center-of-mass energy $\sqrtsnn=5.5$~TeV.

For our exploratory numerical studies, we apply the kinematic cuts provided in Table~\ref{tab:AA-cuts}.
Most notably, the cuts on the minimal $\ptgam$ and $\ptQ$ are asymmetric 
in order to avoid configurations with $\ptgam \simeq \ptQ$
where the NLO cross section is known to become infrared sensitive \cite{Stavreva:2010mw}.
For simplicity, we consider events at mid-rapidity in small bins around $y_\gamma=0$ and $y_Q=0$.
Moreover we impose a cut on the azimuthal angle between the photon and heavy quark jet, $\Delta \phi_{\gamma Q} >3\pi/4$
in order to ensure that the photon and the heavy quark are produced in different hemispheres.
Once data are available with sufficient statistics these results could be updated with the appropriate experimental cuts.
\begin{table}[h]
\begin{center}
\begin{tabular}{c|c|c|c|c}
& $\pt$ & Rapidity & Photon isolation & Jet radius  
\\
\hline \hline 
Photon & $p_{T,\gamma}^{min} = 20\ \GeV$ & $|y_\gamma|<0.2$ & $R=0.4$, $\epsilon <0.1 E_\gamma$ & ---
\\
Heavy quark jet & $p_{T,Q}^{min} = 12\ \GeV$ & $|y_Q|<0.2$ &  --- & $R= 0.4$
\\
\end{tabular}
\caption{\label{tab:AA-cuts} Phase space cuts used for the theoretical predictions in our study.}
\end{center}
\end{table}

\subsection{Total cross sections and event rates}

In this section we compute the NLO cross sections for ${\rm p}+{\rm p} \to \gamma+c +\X$ and ${\rm p}+{\rm p} \to \gamma+b+\X$ 
taking into account the cuts on the phase space listed in Table~\ref{tab:AA-cuts}.
The corresponding cross sections for Pb--Pb collisions
are obtained by scaling the ones for p--p collisions with $A^2$ where $A=207$ is the
atomic number of lead. 
In Table~\ref{tab:xsec}, we present results for the cross section in vacuum 
and for two media with parameters 
$R=10^5$, $\omega_c=50\ \GeV$
and  
$R=10^5$, $\omega_c=100\ \GeV$, respectively.
The expected number of events per year are given
in the third column assuming a yearly luminosity of
${\cal L}_{\rm PbPb}^{\rm year}=0.5$~nb$^{-1}$ 
for Pb--Pb collisions at the LHC~\cite{Arleo:2003gn}.
As can be seen, the event numbers are sufficiently large to measure the cross section
for both, $\gamma+c$ and $\gamma+b$ production where the $\gamma + b$ cross section
is roughly a factor of  7 smaller.
The smaller cross sections in medium are due to the fact that fewer events
pass the cut $\ptQ > p_{T,Q}^{min} $ when energy loss is present. 
With sufficiently large statistics, it could also be interesting to 
analyze the cross section as a function of $p_{T,Q}^{min}$.
The $\gamma+c$ cross section clearly depends on $\omega_c$.
The suppression factor of the total cross section is given by $\raa = 0.87$ ($\raa=0.74$) 
for $\omega_c=50\ \GeV$ ($\omega_c=100\ \GeV$). 
Conversely, the dependence of the $\gamma + b$ cross section on $\omega_c$ 
is much more modest
with $\raa = 0.95$ ($\raa = 0.93$)
for $\omega_c=50\ \GeV$ ($\omega_c=100\ \GeV$). 
It is noteworthy that the suppression effect could be further enhanced
by optimizing the values for the cuts on $\ptgam$ and $\ptQ$.
In particular, the quenching could be enhanced
(without reducing the cross section too much)
by imposing an upper integration bound $\ptgam <p_{T,\gamma}^{max} \lesssim 50\ \GeV$ 
as can be inferred from Fig.~\ref{fig:sigpta}. 

\begin{table}[t]
\begin{center}
\begin{tabular}{c||c|c|c}
& $\sigma^{pp}_{\gamma+Q}$ [pb] & $\sigma^{Pb Pb}_{\gamma+Q}$ [nb] & $N^{Pb \ Pb}_{\gamma+Q}$ 
\\
\hline \hline 
$\gamma+c$ (noEL) & $112.5$ & 4820 & 2410
\\
$\gamma+c$ ($\omega_c=50\ \GeV$) & $98$ & 4200 & 2100
\\
$\gamma+c$ ($\omega_c=100\ \GeV$) & $83$ & 3556 & 1778
\\
$\gamma+b$ (noEL) & $15.5$ & 664 & 332
\\
$\gamma+b$ ($\omega_c=50\ \GeV$) & $14.7$ & 630 & 315
\\
$\gamma+b$ ($\omega_c=100\ \GeV$) & $14.4$ & 617 & 308
\\
\end{tabular}
\caption{\label{tab:xsec}{Total integrated cross section and number of events 
per year for $\gamma +Q$ production in Pb--Pb collisions at the LHC.}}
\end{center}
\end{table}

The numbers in Table~\protect\ref{tab:xsec} can be used to obtain rough estimates
of cross sections and event rates when other phase space cuts are employed
or detection efficiencies for the photon and the heavy quark are taken into account:
(i) Assuming an approximately flat rapidity dependence, the cross section will roughly scale with the size
of the rapidity bins of the photon and the heavy quark jet. We checked that using $|y_\gamma| < 2$ and
$|y_Q|<2$ leads to a cross section roughly 60 times larger than the
one quoted in Table~\protect\ref{tab:xsec};
(ii) Assuming a power-law behavior for the $\ptgam$ and $\ptQ$ dependence (see Sec.~\ref{sec:raa})
the cross sections will depend strongly on the values of $p_{T,\gamma}^{min}$ and $p_{T,Q}^{min}$.
Results for $p_{T,\gamma}^{min}$ and $p_{T,Q}^{min} $ different from the ones in Table~\ref{tab:AA-cuts}
can be estimated by multiplying the cross sections/event numbers
with a factor $(12\ \GeV/p_{T,Q}^{min})^n \times  (20\ \GeV/p_{T,\gamma}^{min})^n$ with $n\simeq 4$.

While a more precise estimate of the expected event numbers clearly 
depends on the details of the experimental acceptances and efficiencies
our results indicate that the number of $\gamma+c$ events produced in a year 
of Pb--Pb collisions at the LHC will be substantial and that
also a measurement of the $\gamma+b$ cross section will be feasible
before the projected upgrade of the luminosity during the second long shutdown
of the LHC where an instantaneous luminosity 
${\cal L}_{\rm PbPb}^{\rm  inst}=6\times 10^{27}$~cm$^{-2}$~s$^{-1}$ is expected
\cite{alice-upgrade}
corresponding to a yearly ($t=10^6~s$) luminosity of ${\cal L}_{\rm PbPb}^{\rm year}=6$~nb$^{-1}$. 

\subsection{Single-inclusive transverse momentum spectra}
\subsubsection{Differential cross sections}
\label{sec:cross sections}

\begin{figure}
\begin{center}
\subfigure[]{
\label{fig:sigpta}
\includegraphics[angle=0,scale=0.35]{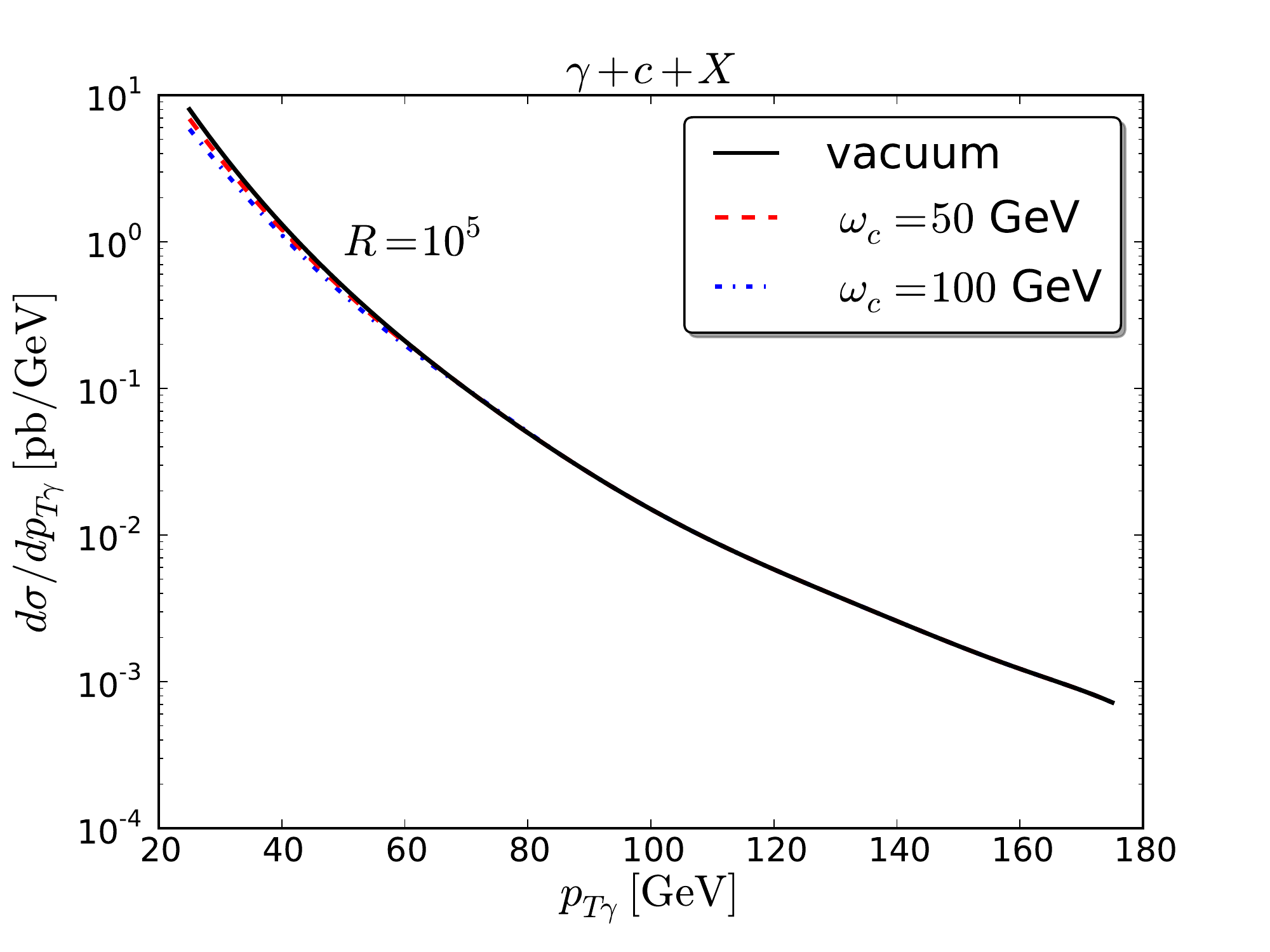}}
\subfigure[]{
\label{fig:sigptb}
\includegraphics[angle=0,scale=0.35]{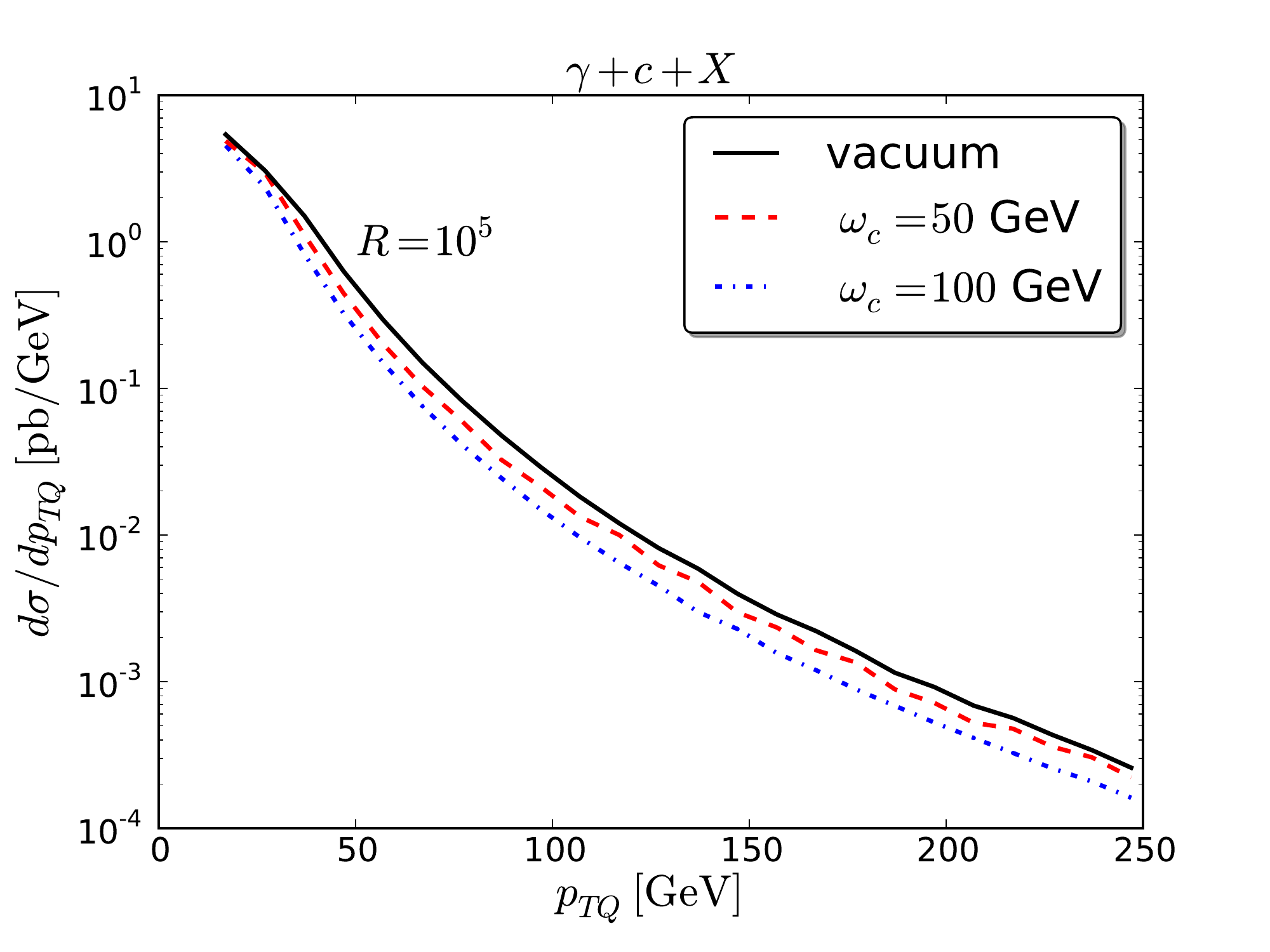}}
\caption{Next-to-leading order differential cross section 
a) $\dd\sigma/\dd\ptgam$  as  a function of $\ptgam$ and 
b) $\dd\sigma/\dd\ptQ$ as a function of $\ptQ$  in pb/GeV.
Shown are results for the vacuum (black solid lines) and for different
media with $R=10^5$ and $\omega_c=50$ (red dashed lines) and $100\ \GeV$ (blue dash-dotted lines).}
\label{fig:sigpt}
\end{center}
\end{figure}

Before turning to the nuclear production ratios, we first present the predictions for the 
single $\pt$ spectra in p--p and Pb--Pb collisions.
In Fig.~\ref{fig:sigpt} the NLO differential cross sections
$\dd \sigma/\dd\ptgam$  (left) 
and $\dd \sigma/\dd\ptQ$ (right)
for $\gamma+c$ production are shown 
for the vacuum and for different media ($\omega_c=50$ and $100\ \GeV$, with $R=10^5$). 
As discussed in Sec.~\ref{sec:framework}, in our framework the photon is not affected by the medium.
However, as can be seen in Fig.~\ref{fig:sigpta}, at small $\ptgam$, the $\ptgam$ spectrum in medium is reduced below
the vacuum spectrum.  The reason is that fewer events pass  the cut on the  heavy quark jet transverse momentum ($\ptQ > 12\ \GeV$), 
if heavy quark energy loss is present. 
This is a physical effect which scales with the heavy quark energy loss at small $\ptgam$: at  $\omega_c = 50\ \GeV$ ($\omega_c = 100\ \GeV$) we 
observe a $15\%$ ($30\%$) reduction.  Conversely $R_{AA}^{\ptgam}$  is
unity at values away from the cut, 
say for $\ptgam \gtrsim 70\ \GeV$, since for these momenta most of the
heavy quark jets pass the cuts as $\ptQmed \gg \ptQmed^{\rm cut}$.   
 
The $\ptQ$ spectra are much more affected by the medium over the entire $p_T$ range, as
is clearly visible in Fig.~\ref{fig:sigptb}.
In this case, the quenching is not related to the presence of the heavy quark threshold
required in the calculation but rather due to the shift of the jet momentum, see Eq.~\eq{eq:shiftQ}, in heavy-ion collisions. 
At  $\ptQ \lesssim 50\ \GeV$ the heavy quark mass mildens the energy loss effects, leading to a small difference
between the vacuum cross section (black solid line) and the in medium cross sections (red dashed line, blue dash-dotted line).  
This will be further discussed in the next section where quenching factors are presented. 

\begin{figure}
\begin{center}
\subfigure[]{
\label{fig:sigpt2a}
\includegraphics[angle=0,scale=0.35]{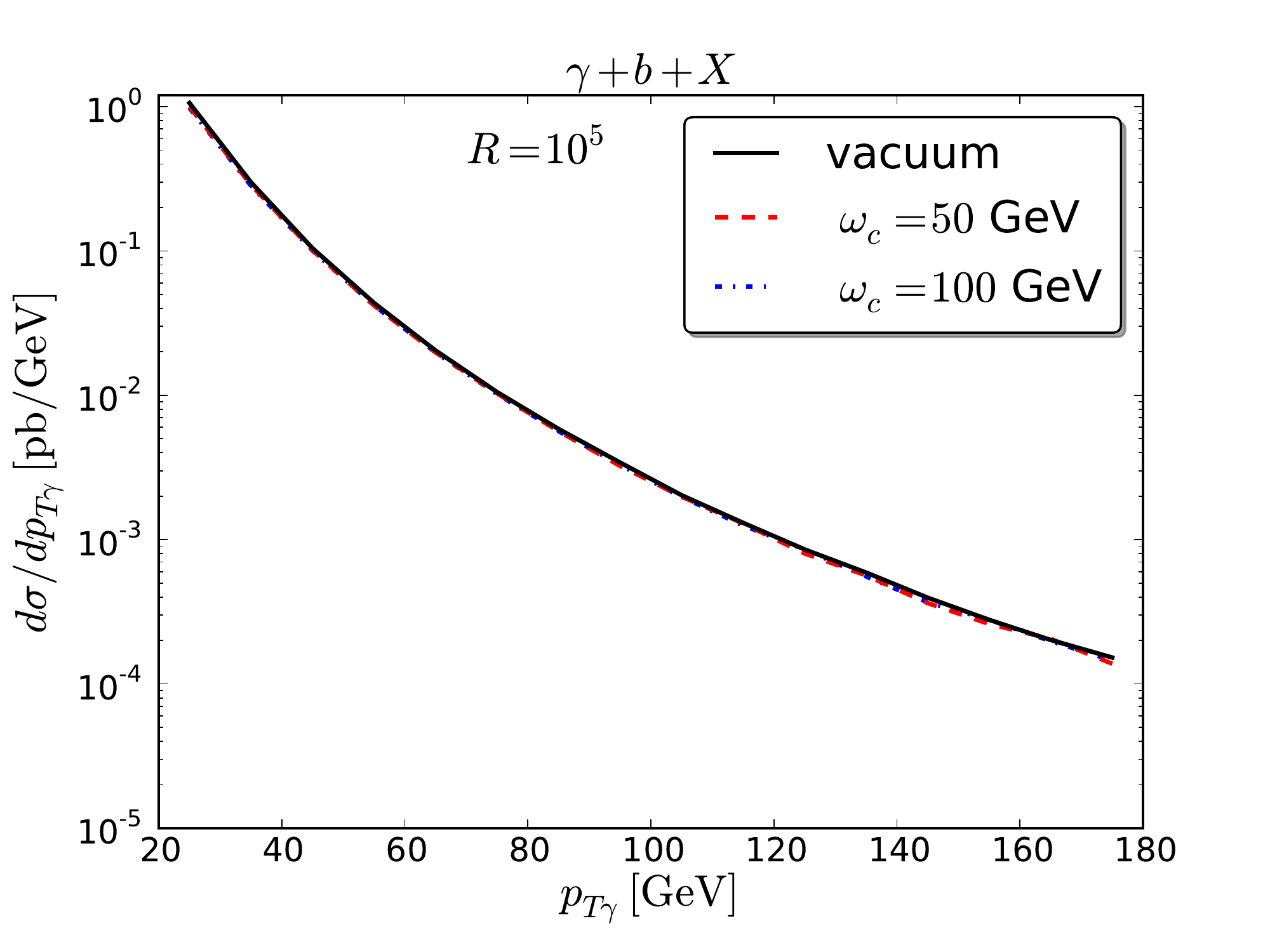}}
\subfigure[]{
\label{fig:sigpt2b}
\includegraphics[angle=0,scale=0.35]{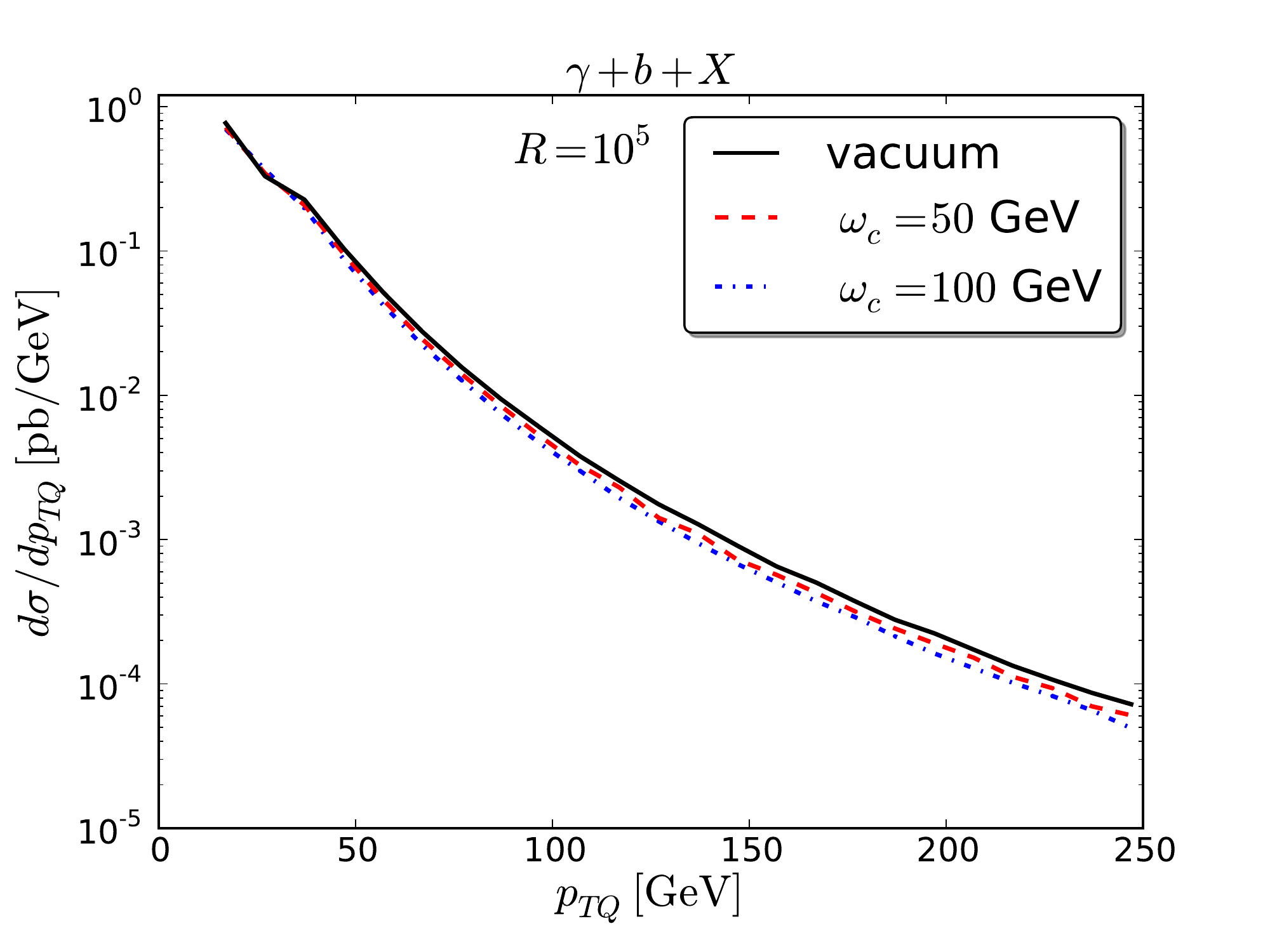}}
\caption{Same as in Fig.~\protect\ref{fig:sigpt} for the $\gamma + b$ case.}
\label{fig:sigpt2}
\end{center}
\end{figure}

In Fig.~\ref {fig:sigpt2} we present the same observables, but for $\gamma+b$ production.  As is visible  
the same trends are present as in  Fig.~\ref {fig:sigpt}, which are however clearly reduced in size compared to the $\gamma+c$ case. 

\subsubsection{Quenching factors $\raa$}
\label{sec:raa}
We consider here  the quenching factor $\raa^Q(\ptQ)$ of the heavy-quark jet. The goal is twofold.
Firstly, as mentioned in Section~\ref{sec:QW}, the aim is to show the sensitivity 
of the observables involving a photon and heavy quark jet in the final state on 
parton energy losses entering via different (massless and massive) quenching weights.
Secondly, we want to explore the effects of the NLO QCD corrections on the quenching factor $\raa^Q(\ptQ)$.

\begin{figure}[thbp]
\begin{center}
\subfigure[]{
\label{fig:RAAQa}
\includegraphics[angle=0,scale=0.35]{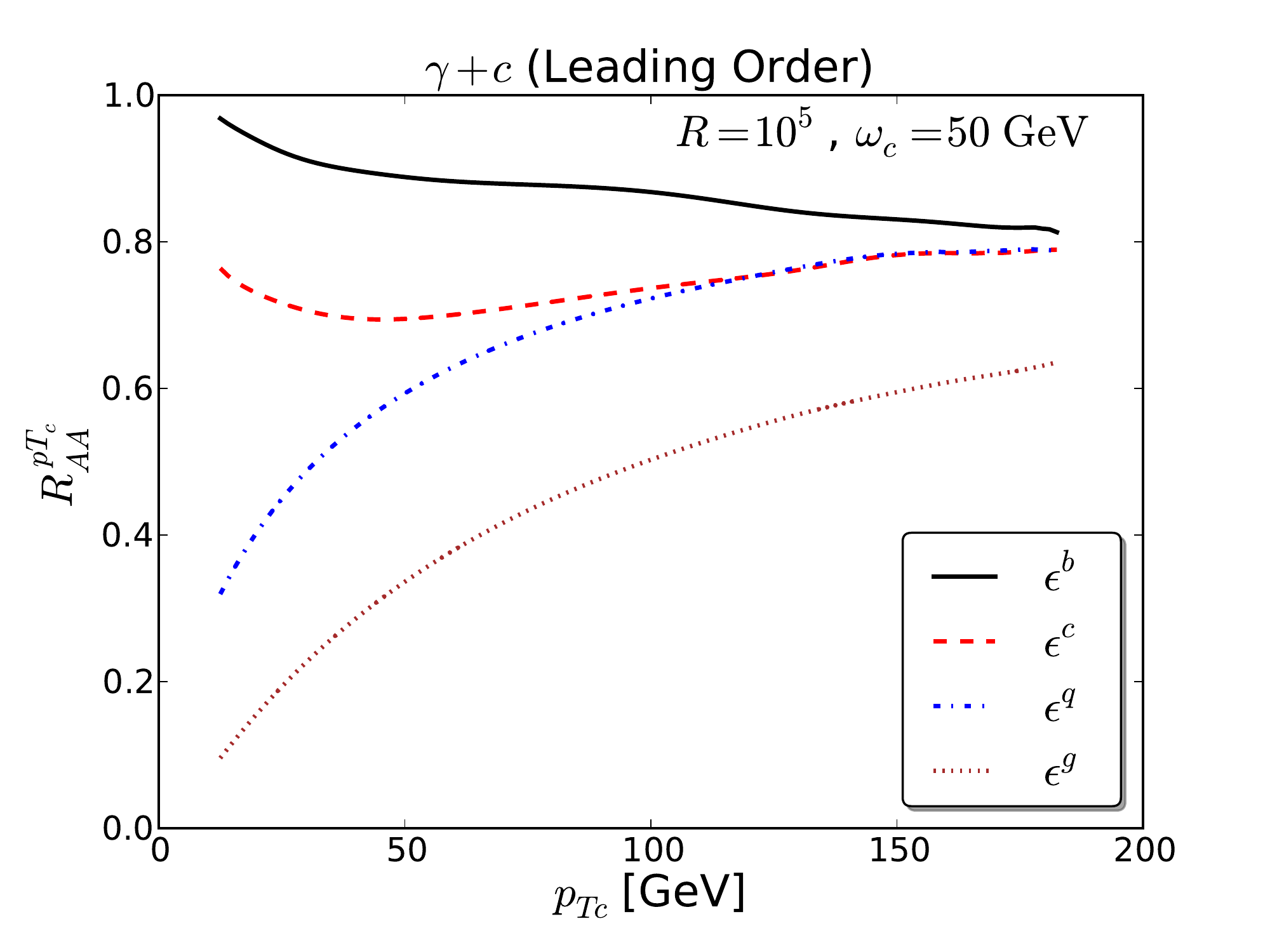}}
\subfigure[]{
\label{fig:RAAQb}
\includegraphics[angle=0,scale=0.35]{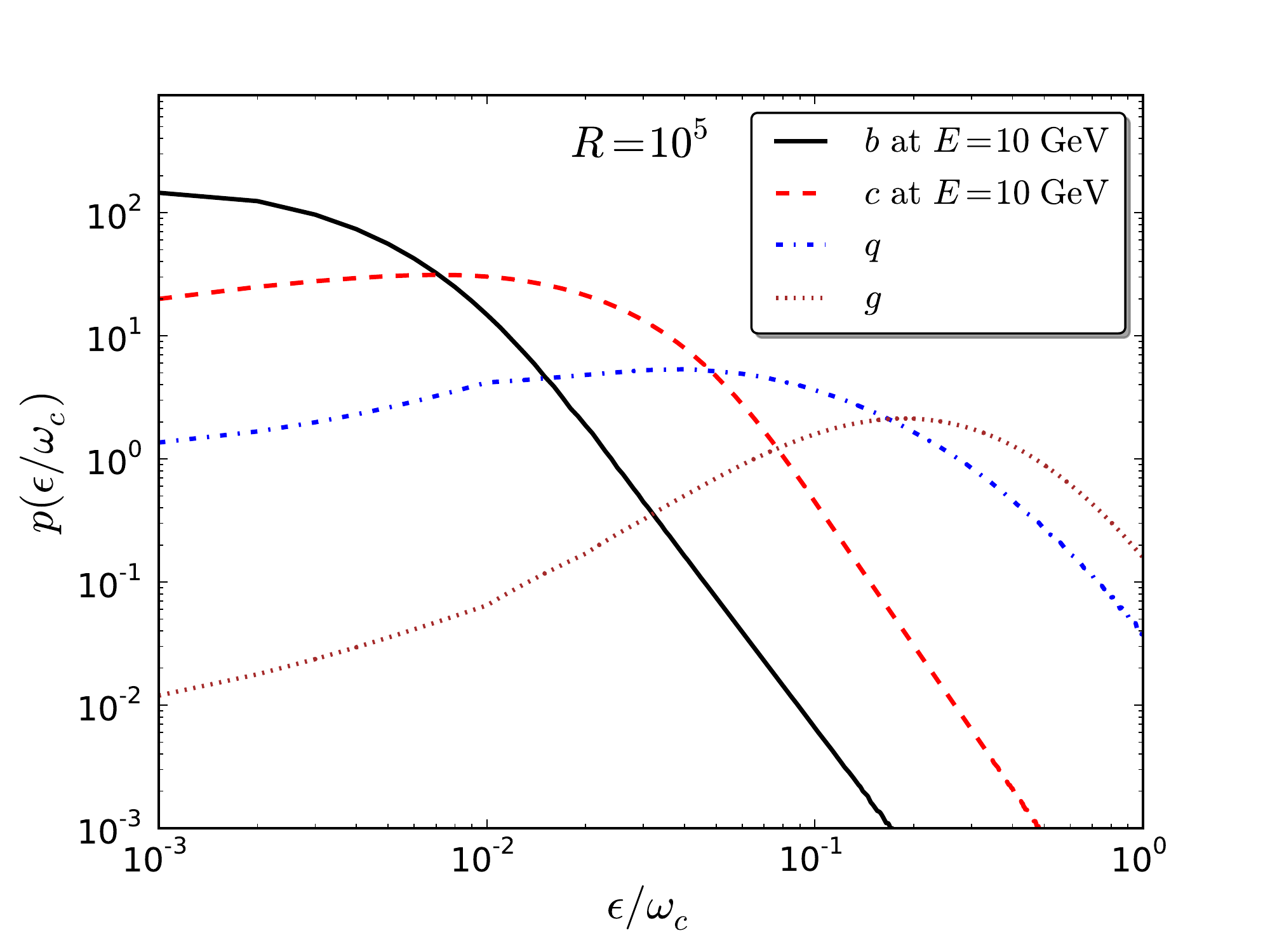}}
\caption{a) Suppression factors $\raa$ in leading order versus transverse charm momentum, $p_{Tc}$, applying 
energy loss according to the quenching weights shown in the right figure; 
b) Quenching weights for energy loss corresponding to gluons (dotted, brown line),
light quarks (dash-dotted, blue line), charm quarks (dashed, red line), and bottom quarks (solid, black line) in 
dependence of the energy loss $\DeltaE$ scaled to $\omega_c$.
}
\label{fig:RAAQ}
\end{center}
\end{figure}

As an illustration of the first point, the quenching of the charm jet
$\pt$ spectrum in $\gamma+c$ production has been computed 
as a function of $p_{Tc}$ in Fig.~\ref{fig:RAAQa} under various
hypothesis regarding the energy loss of charm quarks, i.e.\ assuming 
that the charm quark suffers the same energy loss as that of a light
quark (blue dash-dotted line) or that of a heavy quark assuming $m_Q =
m_c$ (red dashed line) 
and $m_Q = m_b$ (black solid line) in the 
quenching weights (for completeness the calculation has also been
carried out using the quenching weight of a propagating gluon, brown dotted line). 
The calculation is performed at leading order direct, i.e. according to the Compton subprocess only. 
The quenching factors computed using these various prescriptions
follow closely the hierarchy of parton energy loss in the BDMPS-Z
quenching weights used for each flavor in the calculation, which are shown in Fig.~\ref{fig:RAAQb}.
In particular, a much stronger (respectively, weaker) charm jet
quenching would be expected should the charm quark lose energy 
like a gluon (respectively, bottom quark). 
Interesting also is the comparison between the predictions assuming
light quark and charm quark energy loss. 
At low transverse momenta, the charm quark energy loss  predicts less
suppression  as compared to the  light quark energy loss, 
which is comparable in size to the suppression of inclusive D meson production  
as recently measured for example in \cite{ALICE:2012ab}. 
At
large momenta on the other hand, $p_{Tc } \gg m_c$, the induced energy loss of a charm
and a light quark become identical, as can be seen in Fig.~\ref{fig:RAAQa} 
(by comparing the dashed and dash-dotted lines).\footnote{Similarly, the solid black line 
merges with the quenching factor for light quarks at even larger $p_{Tc}$.}

It is possible to extract information on the parton energy loss from the suppression factors $\raa$ in Fig.~\ref{fig:RAAQa}, by assuming a power law dependence of the differential cross section $\dd\sigma/\dd\ptQ$ vs. $\ptQ$:
\begin{equation}
\frac{d \sigma^{\rm vac}}{d \ptQ} \propto \ptQ^{-n}
\end{equation}
where the index $n$ increases smoothly from $n\simeq4$ to $n\simeq5.5$ from low to high transverse momenta.
The medium cross section is obtained by shifting the transverse momentum by a \emph{typical} transverse energy loss $\DeltaETguess$ (which 
could have a $\ptQ$ dependence):
\begin{equation}
\frac{\dd \sigma^{\rm med}}{\dd \ptQ^{\rm med}}(\ptQ^{\rm med}) \simeq \frac{\dd \sigma^{\rm vac}}{\dd
  \ptQ}(\ptQ=\ptQ^{\rm med} + \DeltaETguess) \, .
\end{equation}
As a consequence the suppression factor is approximately given by
\begin{equation}
\raa^Q(\ptQ) = \left( \frac{\ptQ}{\ptQ + \DeltaETguess} \right)^n \ ,
\label{Eq:raa}
\end{equation}
leading to
\begin{equation}
\DeltaETguess = \ptQ  \left( (\raa^Q)^{-1/n}  - 1 \right)\, , \, \quad
\DeltaEguess= \DeltaETguess \cosh y_Q\, .
\label{Eq:eps}
\end{equation}

\begin{table}
\begin{center}
\begin{tabular}{c|ccccccc}
 & $\ptQ$ & $R_{AA}$ & $n$  & $\DeltaEguess$ (GeV) &  $\DeltaEguess/\omega_c$ &$\DeltaE^{\rm peak}/\omega_c$ & $\langle\DeltaE^{\rm qw}\rangle/\omega_c$
\\
\hline
$b$ & 30 & 0.90 & 4.0 & 0.8 & 0.016 & 0.008 & 0.022
\\
$c$ & 30 & 0.69 & 4.0 & 2.9 & 0.058 & 0.025 & 0.08
\\
$q$ & 30 & 0.49 & 4.0 & 6.0 & 0.12 & 0.04 & 0.18		
\\
$g$ & 30 & 0.23 & 4.0 & 13.5 & 0.27 & 0.19 & 0.4		
\\
\hline
$b$ & 90 & 0.87 & 5.5 & 2.3 & 0.046 & 0.026 & 0.069
\\
$c$ & 90 & 0.72 & 5.5 & 5.6 & 0.11 & 0.036 & 0.18
\\
$q$ & 90 & 0.70 & 5.5 & 6.0 & 0.12 & 0.04 & 0.18
\\
$g$ & 90 & 0.48 & 5.5 & 13.0 & 0.26 & 0.19 & 0.4
\end{tabular}
\end{center}
\caption{Extracted energy loss (column 4) from the nuclear suppression factor $R_{AA}$ (column 2) in 
Fig.~\protect \ref{fig:RAAQa} using Eq.~\protect \eqref{Eq:eps} for bottom, charm,  light quark, and 
gluon energy loss, at two different $\ptQ$ values (column 1).  For comparison the peak energy loss 
(column 6) and the mean energy loss (column 7) scaled to $\omega_c = 50 \ \GeV$ are shown as well,  
see text for details.}
\label{tab:tab1}
\end{table}

Using Eq.~\eqref{Eq:eps} and the values for $\raa$ in Fig.~\ref{fig:RAAQa} at $\ptQ=30 \ \GeV$ (taking $n=4$) and  
$\ptQ=90 \ \GeV$ ($n=5.5$) we obtain the energy loss values
$\DeltaEguess$ presented in column four\footnote{Note that $\DeltaEguess \simeq \DeltaETguess$ for central
rapidities $|y|<0.2$.} 
of Table~\ref {tab:tab1} (in the fifth column the value is scaled by $\omega_c =50\ \GeV$).  
We compare this to the energy loss corresponding to the peak of the quenching weights for bottom, 
charm, light quark and gluons in column 6.  Finally in the seventh column we show the energy loss corresponding to the 
mean of these quenching weights.

It is interesting to note that the ratio of the energy loss for gluon and 
light quarks ($13.5/6.0$ or $13.0/6.0$) is very close to the ratio of  the color factors $C_A/C_F = 9/4$.
As can be seen  the energy loss for light quarks and gluons is independent 
of the chosen $\ptQ$ value, whereas the heavy quarks lose less energy
at small $\ptQ$ due to the mass effects.  
Furthermore  at $\ptQ=90 \ \GeV$ the 
charm quark energy loss is already almost identical to the light quark one.  
As can be seen the extracted values for the energy loss (in column five) 
are systematically larger (by a factor varying between 1.4 and 3.0)  than the most likely (``peak'') energy 
loss according to the quenching weights (in column 6)
and systematically smaller (by a factor varying between 1.2 and 1.6) than the average energy loss.

\begin{figure}
\begin{center}
\subfigure[]{
\label{fig:RAANLOa}
\includegraphics[angle=0,scale=0.35]{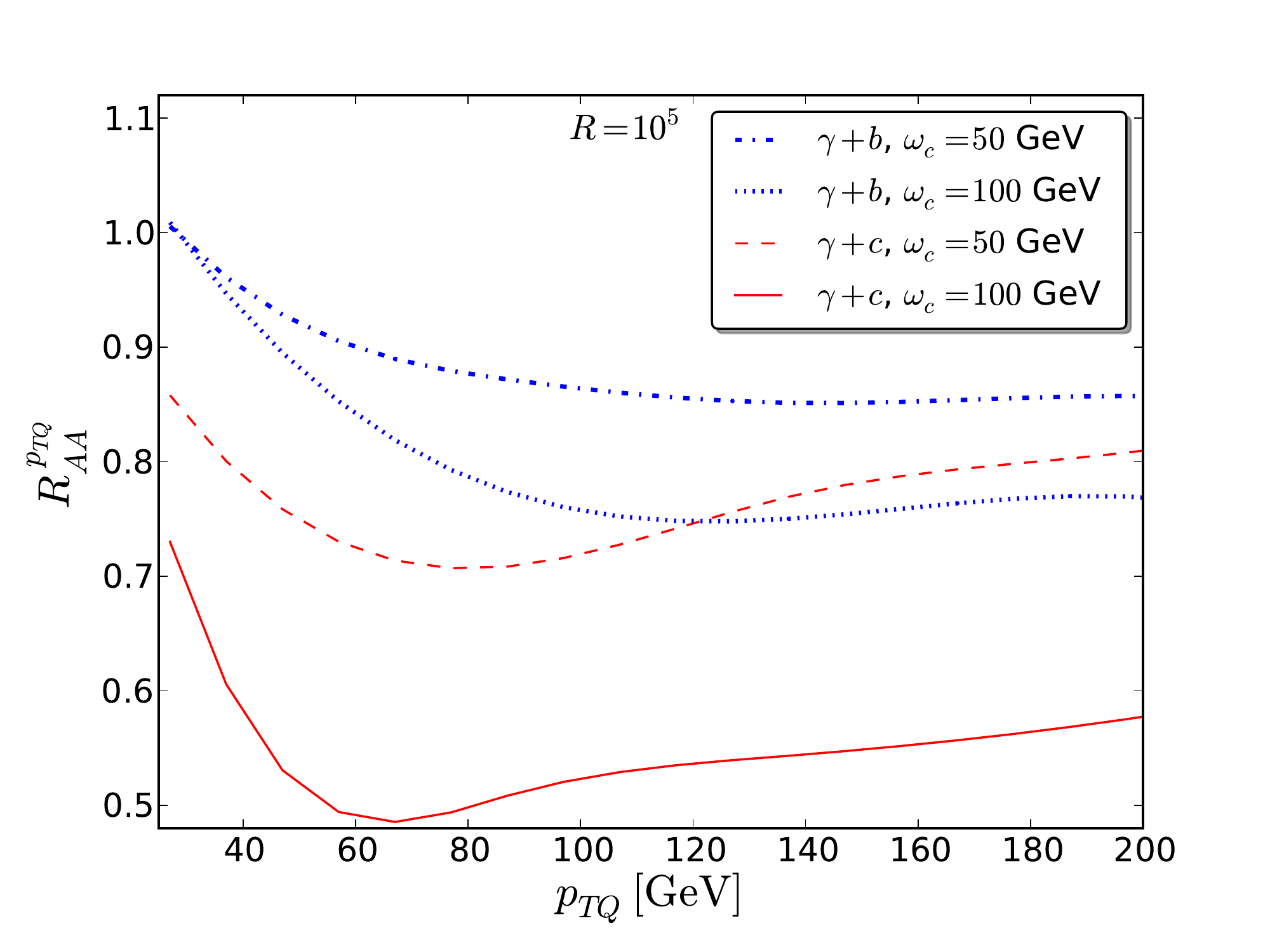}}
\subfigure[]{
\label{fig:RAANLOb}
\includegraphics[angle=0,scale=0.35]{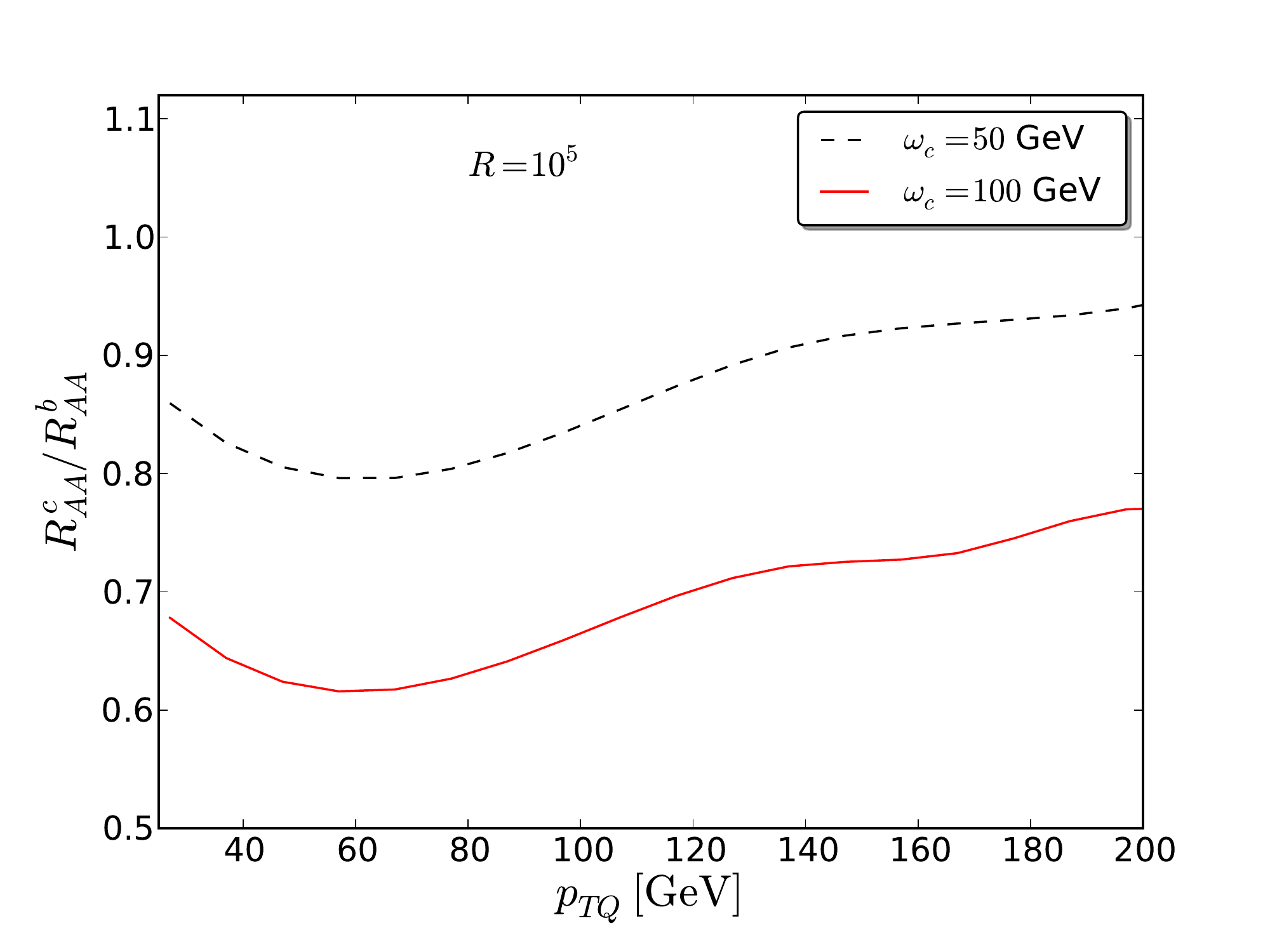}}
\caption{a) $\raa^{\ptQ}$ for $\gamma+c$ and $\gamma+b$ production at NLO QCD. 
b) Double ratio $\raa^c/\raa^b$ for $\omega_c=50$ (dashed line) and $100\ \GeV$ (solid line).}
\label{fig:RAANLO}
\end{center}
\end{figure}

Now turning to the NLO results, we present predictions for the quenching factors
$\raa$ for both, $\gamma+c$ (red dashed and solid lines) and $\gamma +b$ (blue dotted and dash-dotted lines) 
production, in Fig.~\ref{fig:RAANLOa}. 
It is reassuring to see that the quenching factors at NLO are not very different from the ones obtained at LO
(compare, for example, the red dashed lines in Figs. \ref{fig:RAANLOa} and \ref{fig:RAAQa}).
The corresponding double ratios $\raa^c/\raa^b$ for $\omega_c=50$ (dashed line) and $100\ \GeV$ (solid line)
are shown in Fig.~\ref{fig:RAANLOb}.  Clearly, the double ratio for $\omega_c=100\ \GeV$ is much
smaller than the one for $\omega_c=50\ \GeV$ implying a strong sensitivity to the energy loss of the charm and 
bottom quark jets in the medium.
Interestingly, the two double ratios have a very similar shape for both values of $\omega_c$.

\subsection{Two-particle observables}

\begin{figure}
\begin{center}
\subfigure[]{
\label{fig:qTca}
\includegraphics[angle=0,scale=0.35]{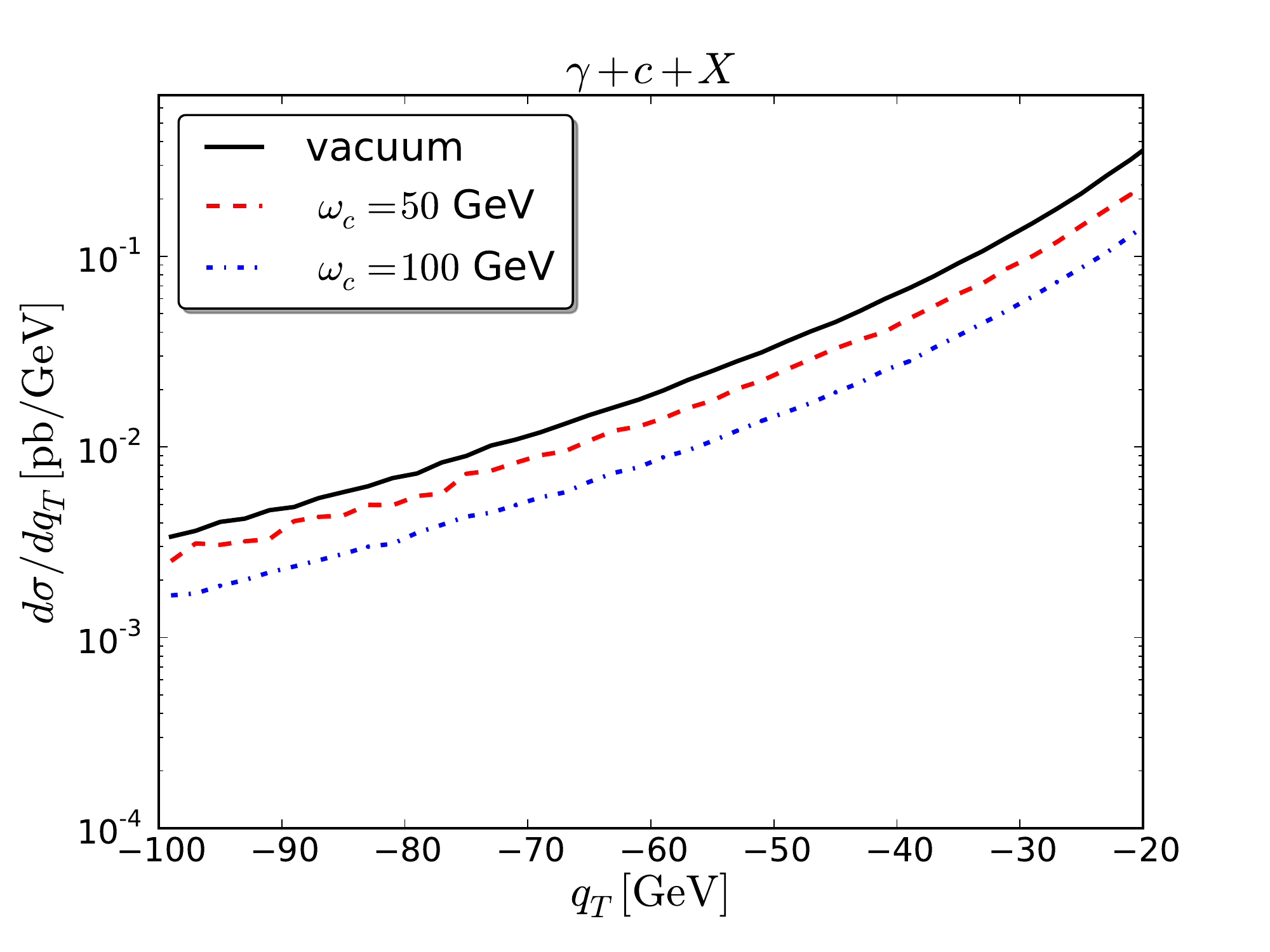}}
\subfigure[]{
\label{fig:qTcb}
\includegraphics[angle=0,scale=0.35]{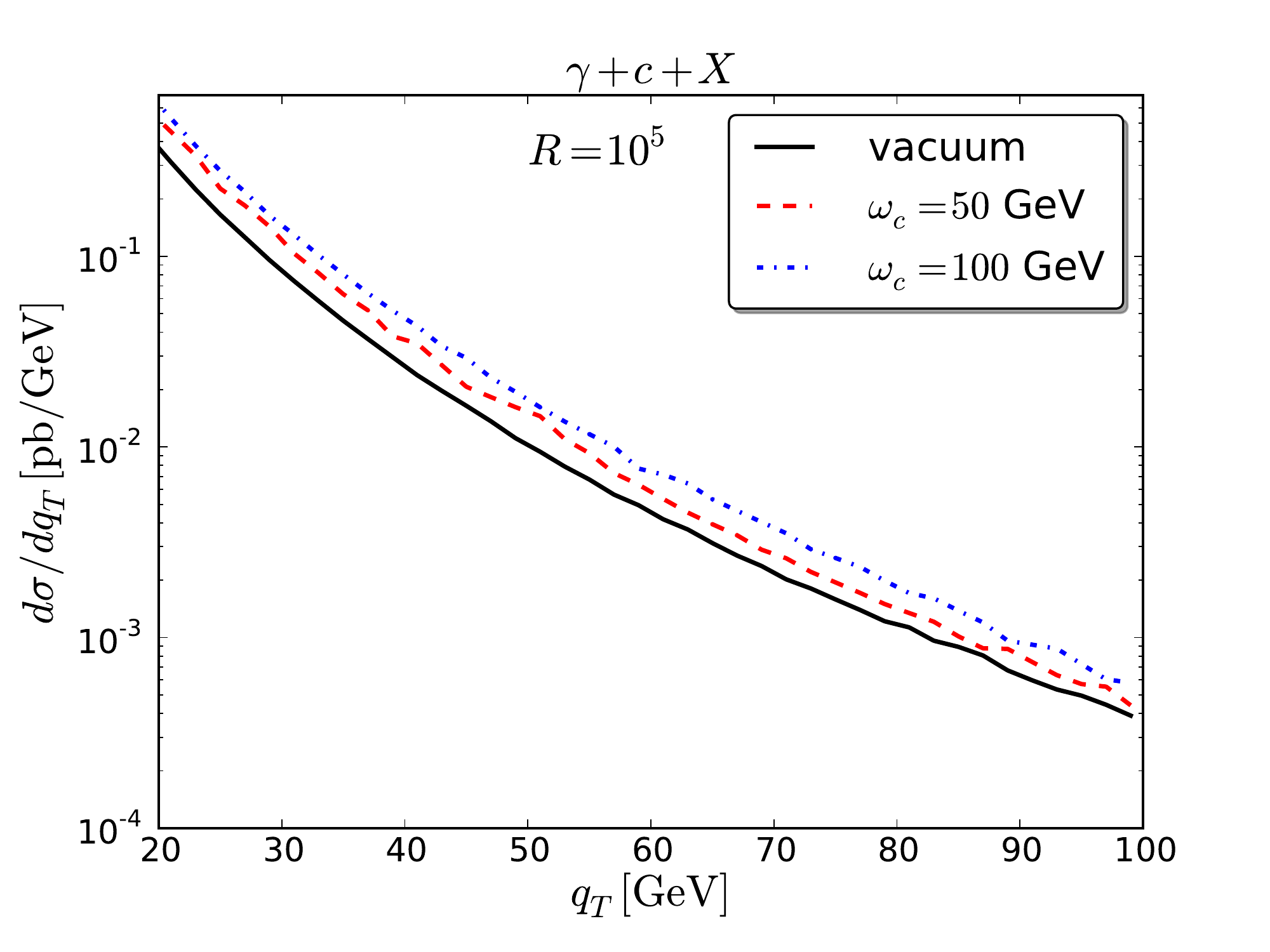}}
\caption{Next-to-leading order differential cross section $d\sigma/dq_T$ for
the production of $\gamma + c$
for a) negative and b) positive values of $q_T$.
Shown are results for the vacuum (black solid lines) and for different
media with $R=10^5$ and $\omega_c=50$ (red dashed lines) and $100\ \GeV$ (blue dash-dotted lines).}
\label{fig:qTc}
\end{center}
\end{figure}

\begin{figure}
\begin{center}
\subfigure[]{
\label{fig:xxx}
\includegraphics[angle=0,scale=0.5]{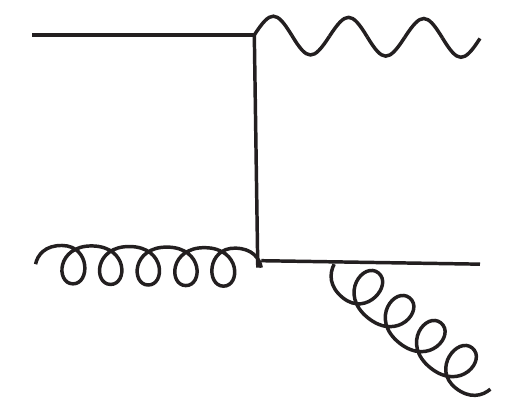}
}
\subfigure[]{
\label{fig:yyy}
\raisebox{14pt}{\includegraphics[angle=0,scale=0.5]{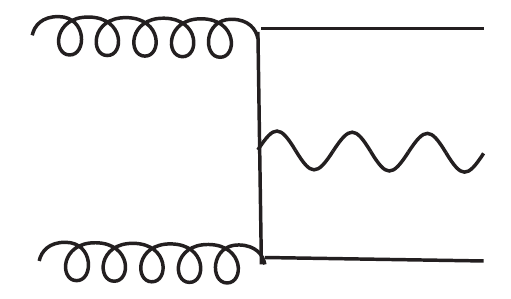}}
}
\caption{Feynman diagrams for the
subprocesses a) $gQ\to g Q \gamma$
and b) $gg \to Q \bar{Q} \gamma$
giving the main contribution to the
$q_T$ distributions in Figs.\ \protect\ref{fig:qTc} and \protect\ref{fig:qTb}
at {\em positive} values of $q_T$.}
\label{fig:diagram1}
\end{center}
\end{figure}

\begin{figure}
\begin{center}
\subfigure[]{
\label{fig:zzz}
\includegraphics[angle=0,scale=0.4]{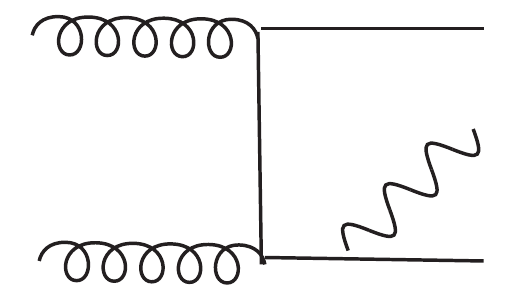}
}
\subfigure[]{
\label{fig:www}
\raisebox{0pt}{\includegraphics[angle=0,scale=0.4]{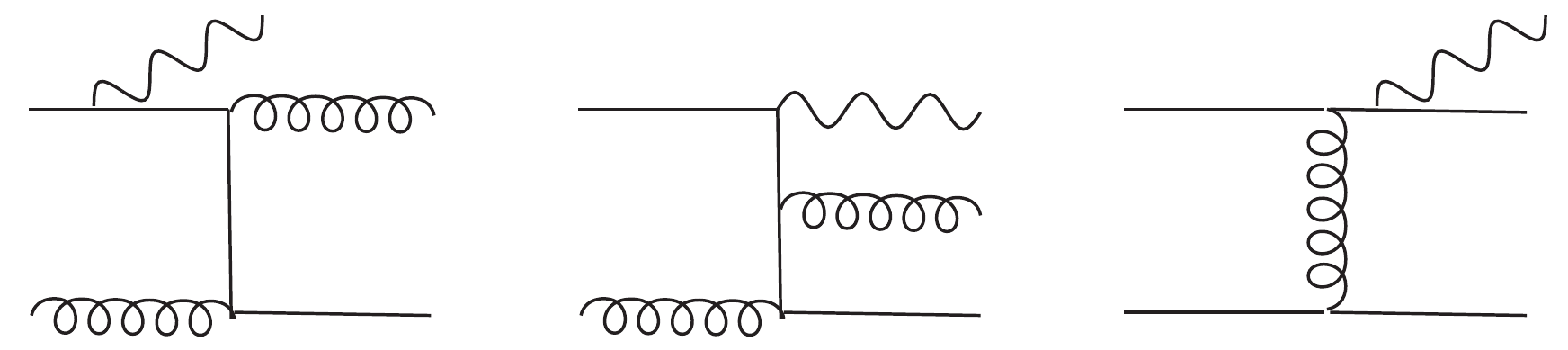}}
}
\caption{Feynman diagrams for the 
subprocesses a) $gg \to Q \bar{Q} \gamma$
and b) $gQ\to gQ\gamma$ and $qQ \to q Q \gamma$
giving the main contribution to the
$q_T$ distributions in Figs.\ \protect\ref{fig:qTc} and \protect\ref{fig:qTb}
at {\em negative} values of $q_T$.}
\label{fig:diagram2}
\end{center}
\end{figure}

\begin{figure}
\begin{center}
\subfigure[]{
\label{fig:AJa}
\includegraphics[angle=0,scale=0.35]{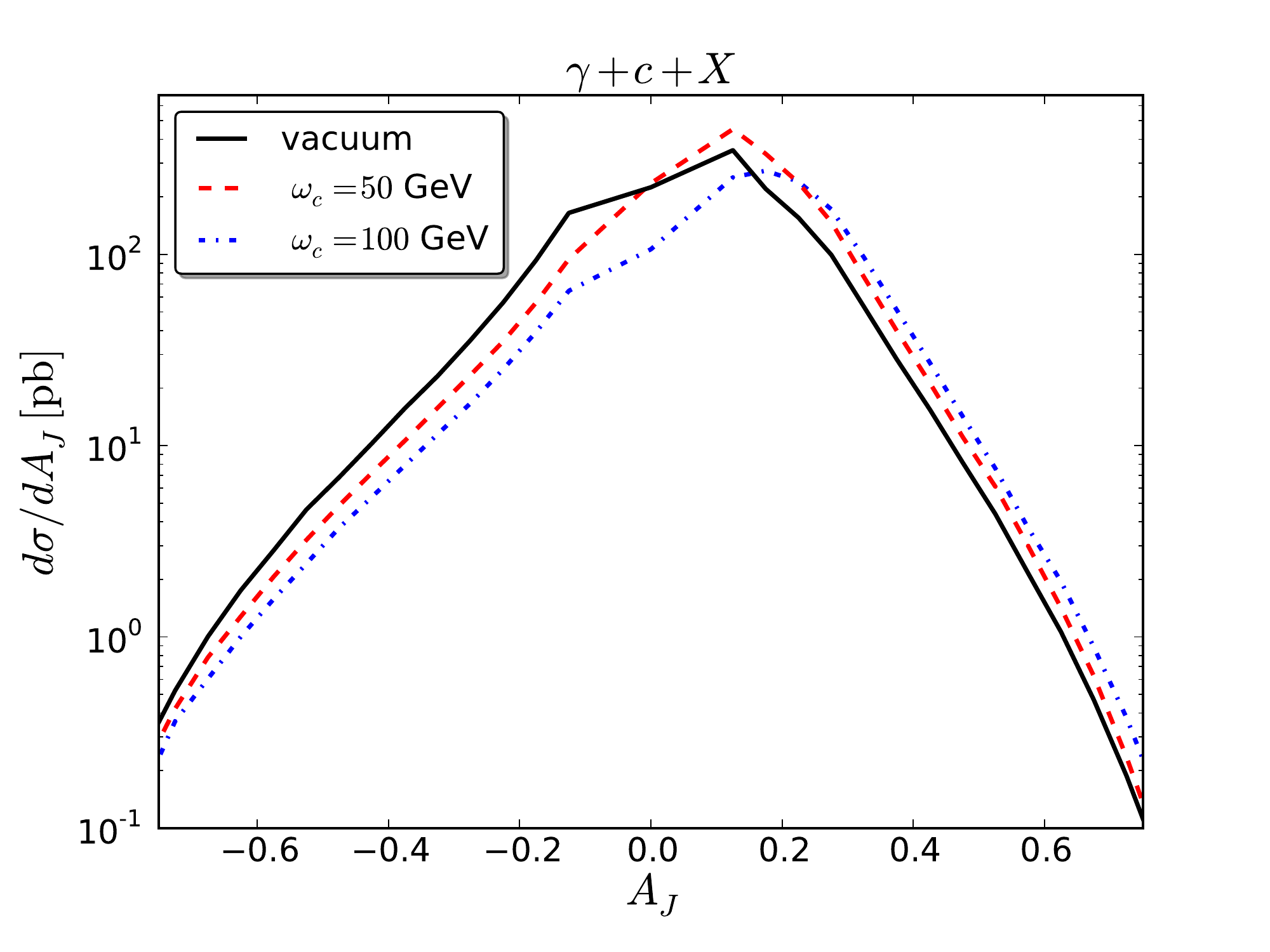}}
\subfigure[]{
\label{fig:AJb}
\includegraphics[angle=0,scale=0.35]{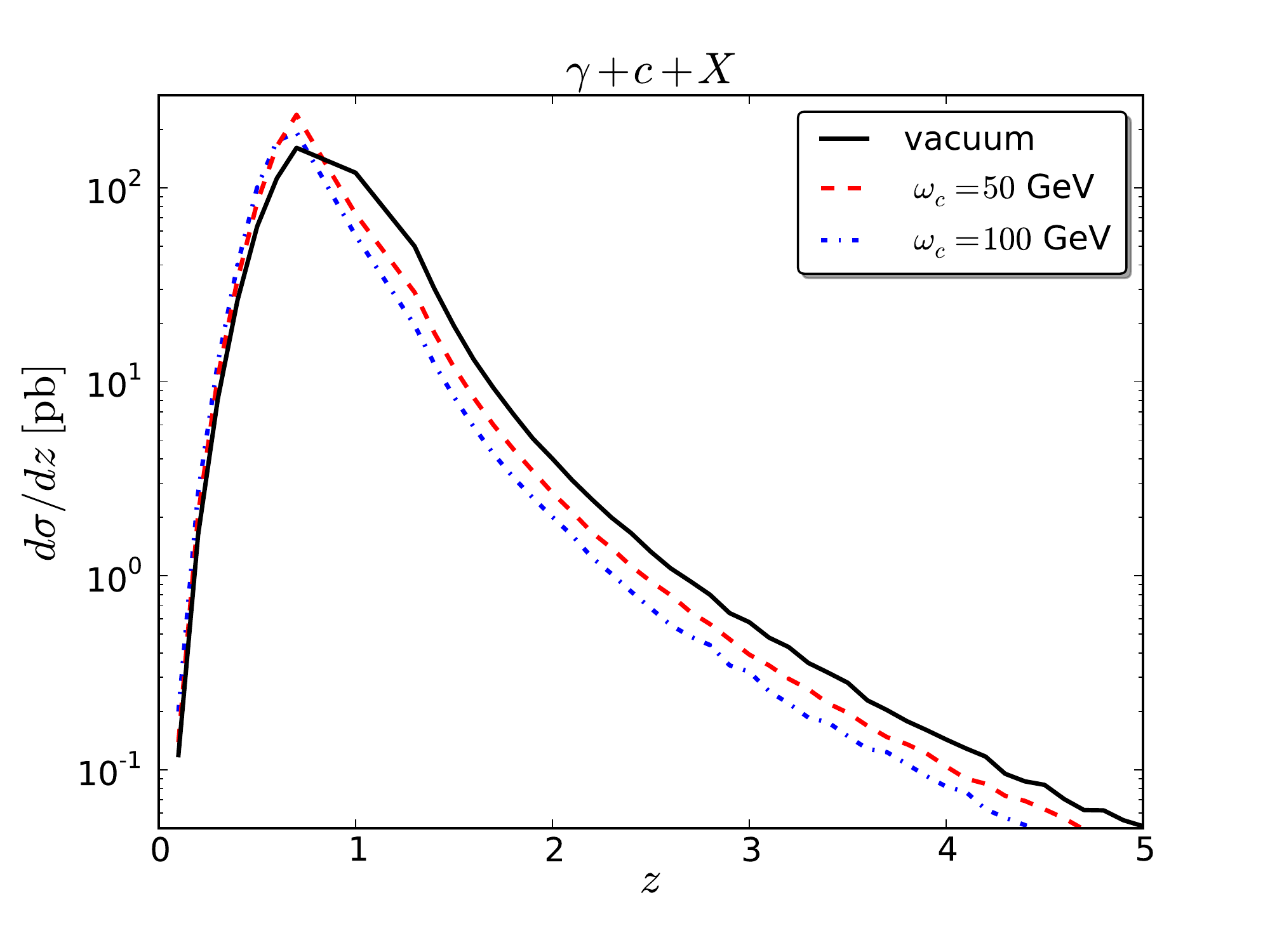}}
\caption{Differential cross section with respect to 
a) the jet asymmetry $A_J$
and 
b) the momentum imbalance $z$
int the $\gamma+c$ production channel.}
\label{fig:AJ}
\end{center}
\end{figure}

We now turn to a discussion of the two-particle observables introduced in Sec.\  \ref{sec:observables}.
We begin with the $q_T$ distribution defined in Eq.~\eqref{eq:qT}  for $\gamma+c$ production 
which is shown in Fig.~\ref{fig:qTca} for negative and in Fig.~\ref{fig:qTcb} for positive values of $q_T$.
We have spared the region $-20\ \GeV \le q_T \le 20\ \GeV$ where $\ptgam \simeq \ptQ$ to avoid problems
with IR-safety\footnote{Note that meaningful results can still be obtained when integrating/binning over 
a sufficiently large region around $q_T=0$ as has been done in Fig.~\protect\ref{fig:AJ} for the observables
$A_J$ (binning around $A_J=0$) and $z$ (binning around $z=1$).}
as discussed in Sec.~\ref{sec:ingredients}.
In addition to the vacuum cross section (black solid line) results are shown for a medium with
$\omega_c=50\ \GeV$ (red dashed line) respectively $\omega_c=100\ \GeV$ (blue dash-dotted line).
As can be seen the medium spectra are right-shifted with respect to the vacuum one 
for both, negative and positive values of $q_T$
(see the discussion in Sec.\  \ref{sec:observables}).
Clearly, the shift increases with increasing $\omega_c$ and is smaller for positive
$q_T$ where it is about $2\ \GeV$ for $\omega_c =50\ \GeV$ and varying between 3.5 and $5\ \GeV$
for $\omega_c=100\ \GeV$.
For negative $q_T$ the shift is roughly $5\ \GeV$ ($11$--$15\ \GeV$) for $\omega_c=50\ \GeV$
($\omega_c=100\ \GeV$).
The overall differences in shape and size of the curves in Figs.~\ref{fig:qTca}  and \ref{fig:qTcb}
can be understood as a consequence of the different subprocesses which contribute (dominantly)
to the cross section for positive and negative $q_T$.
For positive $q_T$, the dominant contribution comes from the subprocess
$gQ\to gQ\gamma$ followed by $gg\to Q\bar{Q}\gamma$ where appropriate
Feynman diagrams are shown in Figs.\  \ref{fig:xxx} and \ref{fig:yyy}, respectively,
where the leading photon is balanced by the $Qg$ pair in Fig.~\ref{fig:xxx}
and by the $Q\bar{Q}$ pair in Fig.~\ref{fig:yyy}.
For negative $q_T$, the dominant contribution comes from the subprocess
$gg\to Q\bar{Q}\gamma$ where in this case 
the relevant Feynman diagram is shown in Fig.~\ref{fig:zzz} for a leading heavy quark jet.\footnote{Note 
that this amplitude is suppressed in the case of positive $q_T$ because we require that the tagged heavy quark be the
one with the largest $p_T$.} 
Other non-negligible subprocesses (for $q_T<0$) are $gQ\to gQ\gamma$ and $qQ\to qQ\gamma$ (see Fig.~\ref{fig:www}
for sample diagrams).

\begin{figure}
\begin{center}
\subfigure[]{
\label{fig:qTba}
\includegraphics[angle=0,scale=0.35]{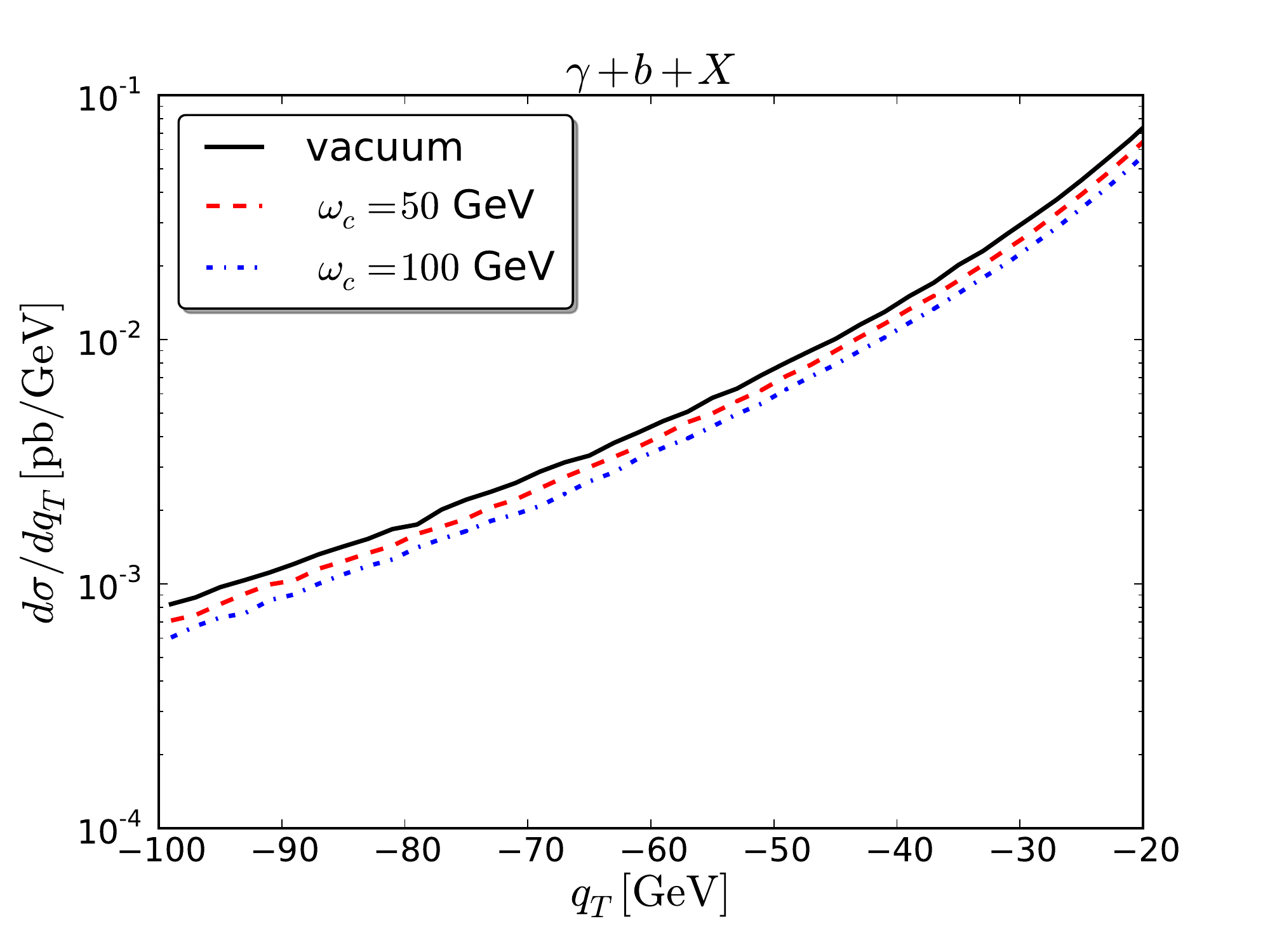}}
\subfigure[]{
\label{fig:qTbb}
\includegraphics[angle=0,scale=0.35]{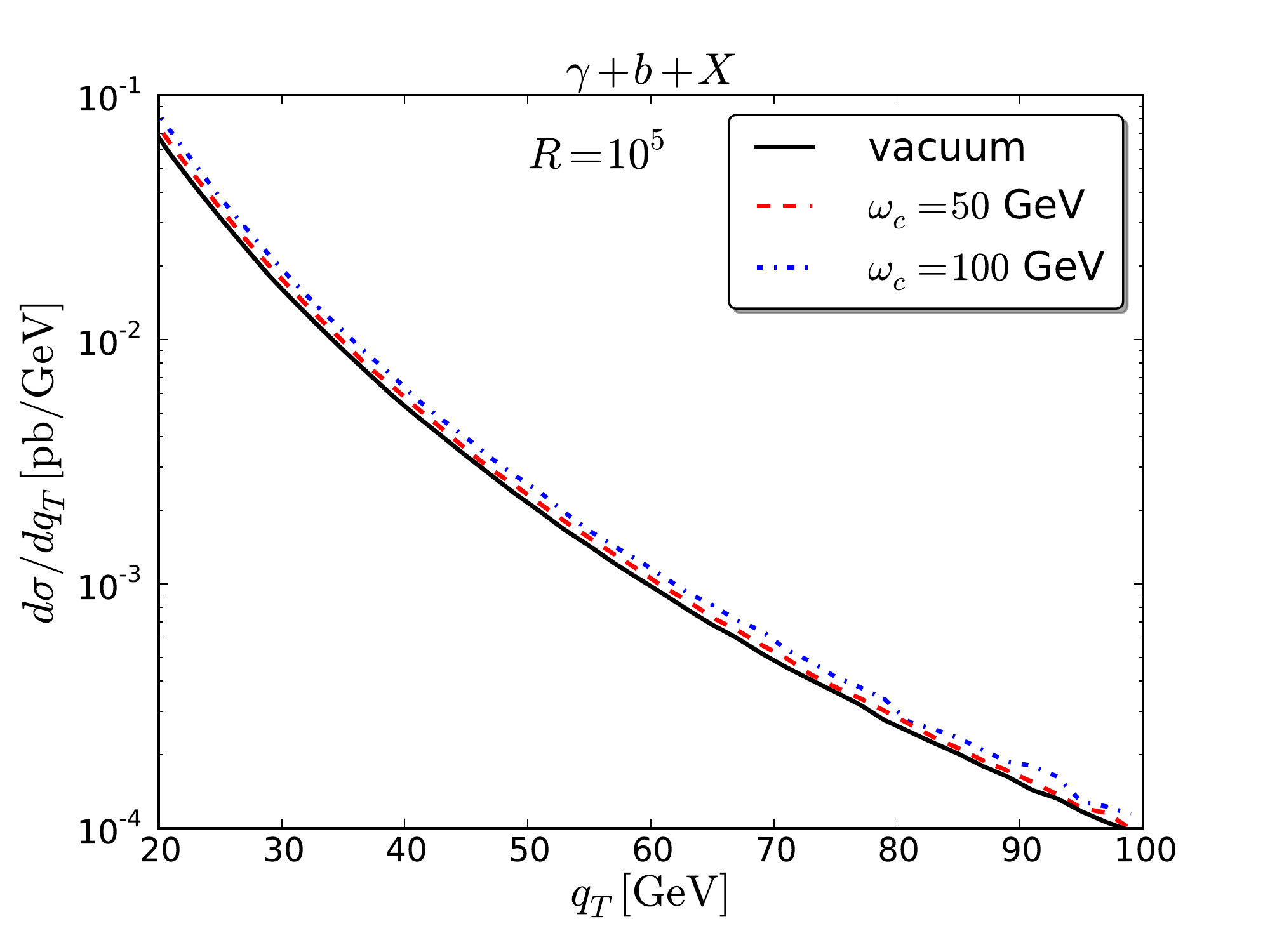}}
\caption{Same as in Fig.~\protect\ref{fig:qTc} for $\gamma+b$ production.}
\label{fig:qTb}
\end{center}
\end{figure}

The larger shift in energy for negative $q_T$ (compared to the positive $q_T$ case) can be understood
in the following way: the photon momentum $\ptgam$ remains fixed close to its minimal value $\ptgam^{\rm min}=20\ \GeV$
whereas $\ptQ$ varies (i.e.\ is larger) in order to satisfy $q_T \simeq  \ptgam^{\rm min} - \ptQ$.
Hence the heavy quark energy loss is less pronounced since mass effects are less important at larger $\ptQ$ values.
Conversely, for $q_T > 0$, the situation is inverted where now
$\ptQ$ remains fixed close to its minimal value $\ptQ^{\rm min} = 12\ \GeV$
whereas $\ptgam$ varies (i.e.\ is larger) in order to satisfy $q_T \simeq  \ptgam - \ptQ^{\rm min}$.
In this case, the heavy quark energy loss will be smaller because the heavy quark mass effects
are still relevant.
In Fig.~\ref{fig:qTb} we present the corresponding results for $\gamma+b$ production.
Again, the same trends can be observed which are, however, reduced in size.

Finally, in Fig.~\ref{fig:AJ} we show results for the jet asymmetry $A_J$ (left) and the
momentum imbalance $z$ (right).
The variable $A_J$ defined in Eq.~\eqref{eq:AJ} is very similar to $q_T$ and the $A_J$ spectrum
shows indeed very similar features as the $q_T$ distribution, in particular the in-medium curves
are shifted to the right, the shift is larger for negative $A_J$, and the distribution is slightly asymmetric
around $A_J=0$.
The $z$-distribution peaks at $z=1$ which corresponds to a configuration where the heavy quark and the photon are
back-to-back (as in LO). With our cut on the angle between the heavy quark jet and the photon momentum
($\theta > 3 \pi/4$) we find $0.7 \times \frac{\ptQ}{\ptgam} \le z=-\frac{\ptQ} {\ptgam} \ \cos \theta \le \frac{\ptQ}{\ptgam}$.
Therefore, roughly, the region $z<1$ ($z>1$) corresponds to $\ptQ < \ptgam$ ($\ptQ > \ptgam$).
In this case, the in-medium spectra are left-shifted and the bigger shifts are visible in the region $z>1$ which
corresponds, as in the previous figures, to the kinematic configuration $\ptQ > \ptgam$.

\section{Conclusions}
\label{sec:conclusions}

An important aspect of jet quenching phenomenology is to understand the mass dependence of parton energy loss. 
While the quenching of single inclusive hadron production might give a clue on this topic, 
the possibilities offered by the LHC experiments allow for investigating more exclusive processes
which have the advantage that a range of interesting observables can be constructed from the momenta 
of the final state particles.

In this paper, we performed an exploratory study of the associated production of a photon and a 
heavy quark jet in heavy-ion collisions, at NLO accuracy. 
On general grounds this is a very promising process
in which the energy loss of the heavy quark jet can be calibrated 
using the prompt photon momentum
since the latter is assumed to be unaffected by the hot and dense QCD medium.
Despite the smaller cross sections as compared to the single inclusive
heavy-quark jet production, the expected rate at the LHC is
sufficiently large to ensure reasonable statistical uncertainties for 
both $\gamma+c$ and $\gamma + b$ production.
In addition to the inclusive $p_T$ spectra of the photon and the heavy quark jet, we have
performed NLO calculations for a range of observables ($q_T$, $A_J$, $z$) in 
order to identify which of these distributions provide useful information 
on the amount of energy loss experienced by the propagating heavy quark.

The inclusive $\ptQ$ and $\ptgam$ distributions are very promising observables. 
Numerical results for  $\dd\sigma/\dd\ptgam$ and $\dd\sigma/\dd\ptQ$ calculated
at NLO QCD were presented.
We find that the $\ptQ$ spectra are affected by the medium over the entire $p_T$ range, whereas the
$\ptgam$ distribution only gets quenched at small $\ptgam$ due to the cut on the jet transverse momentum.
Comparing these two spectra should thus be particularly instructive in order to
disentangle the effects due to the heavy-quark energy loss from the ones due to the
nuclear modification of the parton distribution functions and other cold nuclear matter effects.
Information on the parton energy loss could be obtained either by studying the total cross section as
a function of the cut on the jet transverse momentum or by analyzing the $\ptQ$ distribution.
For the latter we have presented in Fig.~\ref{fig:RAANLOa} the quenching factors $\raa$ for both,
$\gamma+c$ and $\gamma+b$ production at NLO QCD.
Very interesting is also the double ratio $\raa^c/\raa^b$ 
shown in Fig.~\ref{fig:RAANLOb}
of which the {\em shape} (not the normalization) as function of $\ptQ=p_{Tc}=p_{Tb}$ 
turns out to be almost independent of the choice of the parameter $\omega_c$.

The distributions in the two-particle kinematic variables have also
been investigated in detail, paying attention in particular to the
photon--jet pair momentum $\qt$. 
As expected, these distributions are shifted towards larger values in
heavy-ion collisions. 
The comparison of $\qt$ spectra in p--p and Pb--Pb collisions for
$\gamma+c$ and $\gamma+b$ production 
should thus allow for a ``direct'' access to the amount of energy lost
by charm and bottom quarks, respectively. 
Note that such distributions are singular at leading order accuracy
and therefore NLO predictions prove mandatory in order to compute 
such spectra at large (positive and negative) $\qt$ values.

For completeness the distributions in the photon-jet asymmetry, $A_J$,
and momentum imbalance, $z$, have also been determined in the
$\gamma+c$ channel in p--p and Pb--Pb collisions. 
Similar patterns to the ones observed in the $\qt$ distributions are
reported. 
The comparison between the distributions in various kinematic
variables should thus help to determine at a quantitative level the
amount of heavy-quark energy loss from the future high-precision 
measurements through detailed phenomenological studies.

Let us mention some future improvements to be carried out beyond the present study. 
The systematic comparison of $\gamma+c/b$ jet production with $\gamma+$~\emph{inclusive} jet production 
(for which first measurements have been reported recently~\cite{Chatrchyan:2012gt}) should be particularly 
interesting and useful. 
%
In addition, as soon as the nuclear gluon parton distribution is better constrained from data of the
forthcoming p--A run at the LHC, the predictions should be updated using {\em nuclear} PDFs including
systematic studies of the nPDF and scale uncertainties. 
Finally, it would also be important to compare the present calculations with calculations based on other energy loss frameworks 
and different assumptions regarding the modeling of the heavy-quark energy loss in the medium. 
%

\section*{Acknowledgment} 
We are grateful to thank S.\ Peign\'e for useful discussions on parton energy loss.
This work was supported by the CNRS through a PICS research grant and the
Th\'eorie-LHC-France initiative.
This work is funded by ``Agence Nationale de la Recherche'' under grant ANR-PARTONPROP.

\clearpage

\providecommand{\href}[2]{#2}\begingroup\raggedright\endgroup


\end{document}